\documentclass[english]{JHEP3}
\usepackage[T1]{fontenc}
\usepackage[latin1]{inputenc}
\usepackage{amsmath}
\usepackage{graphicx}
\usepackage{amssymb}

\makeatletter

\providecommand{\tabularnewline}{\\}

\makeatletter

\newcommand{\HWPP}{\textsf{Herwig++}}
\makeatletter

\usepackage{cite}

\usepackage[mathscr]{euscript}

\makeatletter

\title{A Positive-Weight Next-to-Leading Order Monte Carlo Simulation of Drell-Yan Vector Boson Production}
\author{Keith Hamilton \\ Centre for Particle Physics and Phenomenology (CP3), \\ Universit\'{e} Catholique de Louvain, \\ Chemin du Cyclotron 2, 1348 Louvain-la-Neuve, Belgium. \\ Email: \email{keith.hamilton@uclouvain.be}}
\author{Peter Richardson\\
   Institute of Particle Physics Phenomenology, Department of Physics, \\ University of Durham,  Durham, DH1 3LE, UK; and\\
Theoretical Physics Group, CERN, CH-1211 Geneva 23, Switzerland.\\Email: \email{peter.richardson@durham.ac.uk}}

\author{Jon Tully \\ Institute of Particle Physics Phenomenology, Department of Physics, \\ University of Durham,  Durham, DH1 3LE, UK. \\ Email: \email{j.m.tully@durham.ac.uk}} 
\preprint{CERN-PH-TH/2008-118 \\ CP3-08-12 \\DCPT/08/80 \\ IPPP/08/40\\MCnet/08/02}

\abstract{
The positive weight next-to-leading-order\,(NLO) matching scheme\,(POWHEG) is applied to
Drell-Yan vector boson production in the {\sf Herwig++} Monte Carlo event generator.
This approach consistently combines the NLO calculation
and parton shower simulation, without the production of negative weight events. 
The simulation includes a full implementation of the truncated shower required
to correctly describe soft emissions in an angular-ordered parton shower, for
the first time. 
The results are compared with Tevatron $W^\pm$ and $Z$ production data and
predictions for the transverse momentum spectrum of gauge bosons at the LHC presented.}

\maketitle

\def\ds#1{\setbox0=\hbox{$#1$}#1\hskip-\wd0\hbox to\wd0{\hss\sl/\/\hss}}
\def\slashb#1{\setbox0=\hbox{$#1$}#1\hskip-\wd0\dimen0=5pt\advance
       \dimen0 by-\ht0\advance\dimen0 by\dp0\lower0.5\dimen0\hbox
         to\wd0{\hss\sl/\/\hss}}

\newcommand{\splusminus}{{\mathchoice%
{\vplusminus\displaystyle}%
{\vplusminus\scriptstyle}%
{\vplusminus\scriptscriptstyle}%
{\vplusminus\scriptscriptstyle}%
}}
\newcommand{\sminusplus}{{\mathchoice%
{\vminusplus\displaystyle}%
{\vminusplus\scriptstyle}%
{\vminusplus\scriptscriptstyle}%
{\vminusplus\scriptscriptstyle}%
}}
\newdimen\hbigcirc
\newdimen\wbigcirc
\newcommand{\vplusminus}[1]{%
\settoheight{\hbigcirc}{$#1\bigcirc$}%
\settowidth{\wbigcirc}{$#1\bigcirc$}%
\makebox[\wbigcirc]{%
\makebox[0pt]{\rule[0.4\hbigcirc]{0.5\wbigcirc}{0.05\hbigcirc}}%
\makebox[0pt]{\rule[0.1\hbigcirc]{0.5\wbigcirc}{0.05\hbigcirc}}%
\makebox[0pt]{\rule[0.1\hbigcirc]{0.05\wbigcirc}{0.6\hbigcirc}}%
\makebox[0pt]{$#1\bigcirc$}}%
}
\newcommand{\vminusplus}[1]{%
\settoheight{\hbigcirc}{$#1\bigcirc$}%
\settowidth{\wbigcirc}{$#1\bigcirc$}%
\makebox[\wbigcirc]{%
\makebox[0pt]{\rule[0.2\hbigcirc]{0.5\wbigcirc}{0.05\hbigcirc}}%
\makebox[0pt]{\rule[0.5\hbigcirc]{0.5\wbigcirc}{0.05\hbigcirc}}%
\makebox[0pt]{\rule[-0.1\hbigcirc]{0.05\wbigcirc}{0.6\hbigcirc}}%
\makebox[0pt]{$#1\bigcirc$}}%
}

\makeatother

\makeatother

\makeatother

\usepackage{babel}
\makeatother
\begin{document}

\section{Introduction}

Monte Carlo simulations traditionally combine leading-order $2\to2$
matrix elements with parton showers which provide resummation of soft
and collinear radiation. This provides a fully exclusive description
of observables which allows evolution to the hadronization scale where
models of the non-perturbative regime can be incorporated. This makes
Monte Carlo simulations an essential tool in experimental analysis
allowing a fully exclusive description of the final state to be compared
directly with experimental results.

Parton shower simulations include the leading-logarithmically,
and an important sub-set of the next-to-leading-logarithmically,
enhanced contributions and therefore underestimate the radiation of high
transverse momentum\,$\left(p_{T}\right)$ partons. The Monte Carlo
description can be improved by matching to higher-order matrix elements
which correctly describe the production of high $p_{T}$ particles.
A number of approaches have been developed to correct the emission
of the hardest parton in an event. In the \textsf{PYTHIA} event generator~\cite{Sjostrand:2006za},
corrections were included for~$e^{+}e^{-}$ annihilation~\cite{Sjostrand:1986hx},
deep inelastic scattering~\cite{Bengtsson:1987rw}, heavy particle
decays~\cite{Norrbin:2000uu} and vector boson production in hadron
collisions~\cite{Miu:1998ju}. In the \textsf{HERWIG} event generator~\cite{Corcella:2000bw,Corcella:2002jc}
corrections were included for ~$e^{+}e^{-}$ annihilation~\cite{Seymour:1991xa},
deep inelastic scattering~\cite{Seymour:1994ti}, top quark decays~\cite{Corcella:1998rs}
and vector boson production~\cite{Corcella:1999gs} in hadron-hadron
collisions following the general method described in~\cite{Seymour:1994df,Seymour:1994we}.%
\footnote{\HWPP\ includes matrix element corrections for ~$e^{+}e^{-}$ annihilation~\cite{Gieseke:2003hm},
top quark decays~\cite{Hamilton:2006ms}, vector boson production~\cite{Gieseke:2006ga}
in hadron-hadron collisions and Higgs production in gluon fusion~\cite{Bahr:2008tx}
using the approach of Refs.~\cite{Seymour:1994df,Seymour:1994we} and the
improved angular-ordered parton shower of Ref.\,\cite{Gieseke:2003rz}.%
} These corrections had to be calculated for each individual process
and only corrected the first or hardest%
\footnote{In \textsf{PYTHIA} the first emission was corrected whereas in \textsf{HERWIG}
any emission which could be the hardest was corrected.%
} emission, in addition the method can only be applied to relatively
simple cases and the leading-order normalisation of the cross section
is retained.

In recent years there have been a number of additional developments
which aim to improve on these results by either providing a description
of the hardest emission together with a next-to-leading order~(NLO)
cross section~\cite{Frixione:2002ik,Frixione:2003ei,Frixione:2005vw,Frixione:2006gn,Frixione:2007zp,Frixione:2008yi,LatundeDada:2007jg,Nason:2004rx,Nason:2006hfa,Frixione:2007nu,Frixione:2007vw,Frixione:2007nw,LatundeDada:2006gx},%
\footnote{There have been other theoretical ideas but only the \textsf{MC@NLO}
and \textsf{POWHEG} methods have led to practical programs whose results
can be compared with experimental data.%
} or the emission of more than one hard parton at leading
order~\cite{Catani:2001cc,Krauss:2002up,Schalicke:2005nv,Lonnblad:2001iq,Mangano:2001xp,Mrenna:2003if}.%
\footnote{A recent comparison of these approaches can be found in \cite{Alwall:2007fs}.%
} These matching prescriptions are complicated because
the regions of phase space filled by the higher-order matrix
elements and the parton shower must be smoothly separated in order
to avoid problems such as double-counting, where the shower and matrix
elements radiate in the same region. In general the major complication
is gaining an analytic understanding of the result of the parton shower
either, to subtract it from the real emission matrix element, as in
the \textsf{MC@NLO} approach~\cite{Frixione:2002ik,Frixione:2003ei,Frixione:2005vw,Frixione:2006gn,Frixione:2007zp,Frixione:2008yi},
or to reweight the real emission matrix elements so they can be merged
with the parton shower in multi-parton matching.

The first successful scheme for matching at NLO was the \textsf{MC@NLO}
approach~\cite{Frixione:2002ik,Frixione:2003ei,Frixione:2005vw,Frixione:2006gn,Frixione:2007zp,Frixione:2008yi}
which has been implemented with the \textsf{HERWIG} event
generator for many processes. The method has two draw backs; first,
it involves the addition of correction terms that are not positive
definite and therefore can result in events with a negative weight
and second, the implementation of the method is heavily dependent
on the details of the parton shower algorithm used by the event generator.

In Ref.\,\cite{Nason:2004rx} a novel method, referred to as \textsf{POWHEG}
(POsitive Weight Hardest Emission Generator), was introduced to achieve
the same aims as \textsf{MC@NLO} while creating only positive weight events and
being independent of the event generator with which it is implemented.
The \textsf{POWHEG} method has been applied to $Z$ pair hadroproduction \cite{Nason:2006hfa},
heavy flavour hadroproduction \cite{Frixione:2007nw} and $e^{+}e^{-}$
annihilation to hadrons \cite{LatundeDada:2006gx}. A general outline
of the ingredients required for \textsf{POWHEG} with two popular NLO subtraction schemes
is given in Ref.\,\cite{Frixione:2007vw}.

The \textsf{POWHEG} shower algorithm involves generating the hardest emission
in $p_{T}$ separately using a Sudakov form factor containing the
full matrix element for the emission of an extra parton and adding
to this vetoed showers, which produces radiation at lower
scales in the shower evolution variable, and a truncated shower, which generates
radiation at higher scales in the shower evolution variable,
than the scale of the highest $p_{T}$ emission. While the \textsf{POWHEG}
scheme is independent of the parton shower algorithm, it does require
the parton shower to be able to produce vetoed and truncated showers.
The ability to perform vetoed showers is present in most modern Monte
Carlo event generators, however some changes are required to enable
them to generate truncated showers. Although the \textsf{POWHEG} approach is
formally correct to the same accuracy as the \textsf{MC@NLO} technique
the two methods differ
in their treatment of sub-leading terms.

In this work the \textsf{POWHEG} approach is applied Drell-Yan vector boson
production with the \HWPP\ \cite{Bahr:2008tx,Bahr:2008pv} event
generator. A full truncated shower is implemented for the first time.%
\footnote{Truncated shower effects were neglected in Refs.\,\cite{Nason:2006hfa,Frixione:2007nw},
while an approximate treatment was used in Ref.\,\cite{LatundeDada:2006gx}
where at most one truncated emission was generated.%
}

The paper is organised as follows. In Sect.~\ref{sec:The-POWHEG-method}
we outline the main features of the \textsf{POWHEG} method. In Sect.~\ref{sec:Next-to-leading-order-cross}
we collect the essential formulae relating to the NLO cross
section, for implementation in the program. In Section~\ref{sec:Implementation}
we give details of the event generation process for the hard
configurations, this is followed by a description of how these
configurations are subsequently reproduced by the \HWPP\  angular-ordered
 parton shower, thereby accounting for colour coherence effects
associated with soft wide angle parton emissions. In Sect.~\ref{sec:Results}
we present the results of our implementation, comparing it to Tevatron
data, and in Sect.~\ref{sec:Conclusion} we give our conclusions.

\section{The POWHEG method\label{sec:The-POWHEG-method}}

In the \textsf{POWHEG} approach~\cite{Nason:2004rx} the NLO differential
cross section for a given \emph{N}-body process can be written as
\begin{equation}
\mathrm{d}\sigma = \overline{B}\left(\Phi_{B}\right)\,\mathrm{d}\Phi_{B}\,\left[\Delta_{\hat{R}}\left(0\right)+\frac{\hat{R}\left(\Phi_{B},\Phi_{R}\right)}{B\left(\Phi_{B}\right)}\,\Delta_{\hat{R}}\left(k_T\left(\Phi_{B},\Phi_{R}\right)\right)\,\mathrm{d}\Phi_{R}\right]\label{eq:powheg_5},
\end{equation}
where $\overline{B}\left(\Phi_{B}\right)$ is defined as
\begin{equation}
\overline{B}\left(\Phi_{B}\right)=B\left(\Phi_{B}\right)+V\left(\Phi_{B}\right)+\int\,\hat{R}\left(\Phi_{B},\Phi_{R}\right)-\sum_{i}\, C_{i}\left(\Phi_{B},\Phi_{R}\right)\,\mathrm{d}\Phi_{R},\label{eq:powheg_6}\end{equation}
$B\left(\Phi_{B}\right)$ is the leading-order contribution,
dependent on the \emph{N}-body phase space variables $\Phi_{B}$, 
the \emph{Born variables}. The regularized virtual term
$V\left(\Phi_{B}\right)$ is a finite contribution
arising from the combination of unresolvable,
real emission and virtual loop contributions.
The remaining terms in square brackets are due to \emph{N}+1-body
real emission processes which depend on both the Born variables
and additional \emph{radiative variables}, $\Phi_{R}$, parametrizing
the emission of the extra parton. 
The real emission term, $\hat{R}\left(\Phi_{B},\Phi_{R}\right)$, 
is given by a sum of parton flux factors multiplied by real emission
matrix elements for each channel contributing to the NLO cross section.
Finally, each term $C_{i}\left(\Phi_{B},\Phi_{R}\right)$
corresponds to a combination of 
\emph{real counterterms}/\emph{counter-event weights}, 
regulating the singularities in 
$\hat{R}\left(\Phi_{B},\Phi_{R}\right)$.
The modified
Sudakov form factor is defined as\begin{equation}
\Delta_{\hat{R}}\left(p_{T}\right)=\exp\left[-\int\mathrm{d}\Phi_{R}\,\frac{\hat{R}\left(\Phi_{B},\Phi_{R}\right)}{B\left(\Phi_{B}\right)}\,\theta\left(k_{T}\left(\Phi_{B},\Phi_{R}\right)-p_{T}\right)\right],\label{eq:powheg_4}\end{equation}
where $k_T\left(\Phi_{B},\Phi_{R}\right)$ is equal to the transverse momentum of the
extra parton in the collinear and soft limits.

In the framework of a conventional parton shower Monte Carlo program
the Sudakov form factor $\Delta_{\tilde{R}}\left(p_{T}\right)$ has
the same form as that in Eq.\,\ref{eq:powheg_4} but with $\tilde{R}\left(\Phi_{B},\Phi_{R}\right)$
replacing $\hat{R}\left(\Phi_{B},\Phi_{R}\right)$, where the former is
typically the sum of the real emission matrix elements approximated
by their soft/collinear limits. This object has an interpretation
as the \emph{probability} that given some initial configuration of
partons resolved at some characteristic scale $p_{T}^{i}$, in \emph{evolving}
to some other resolution scale, $p_{T}$, no further partons are
resolved. 
The parton shower approximation to the NLO differential
cross section is analogously given by
\begin{equation}
\mathrm{d}\tilde{\sigma}=B\left(\Phi_{B}\right)\,\mathrm{d}\Phi_{B}\,\left[\Delta_{\tilde{R}}\left(0\right)+\frac{\tilde{R}\left(\Phi_{B},\Phi_{R}\right)}{B\left(\Phi_{B}\right)}\,\Delta_{\tilde{R}}\left(k_T\left(\Phi_{B},\Phi_{R}\right)\right)\,\mathrm{d}\Phi_{R}\right].
\label{eq:powheg_7}
\end{equation}
The first term in Eq.~\ref{eq:powheg_7} gives a differential distribution
proportional to the leading-order differential cross section, which
is retained in the absence of any parton showering, \emph{i.e.} with
probability $\Delta_{\tilde{R}}\left(0\right)$ (non-radiative events),
while the second term represents the probability of evolving from
the starting scale to scale $p_{T}$ at which point an emission is
generated according to $\left.\tilde{R}\left(\Phi_{B},\Phi_{R}\right)\right|_{p_{T}}$. 
The term in square brackets in Eqs.\,\ref{eq:powheg_5}~and~\ref{eq:powheg_7} is equal to
unity when integrated over the radiative phase space. 

In the \textsf{POWHEG} framework the parton shower is promoted to NLO
 accuracy by demanding that the non-radiative events are distributed
according to the first term in Eq.\,\ref{eq:powheg_5}
and that the hardest (highest $p_{T}$) emission is distributed according
to the second term. Whereas in the conventional simulation an \emph{N}-body
configuration is generated according to $B\left(\Phi_{B}\right)$
and then showered using the Sudakov form factor $\Delta_{\tilde{R}}$,
the \textsf{POWHEG} formalism requires that the \emph{N}-body configuration
is generated according to $\overline{B}\left(\Phi_{B}\right)$ and then showered with
the modified Sudakov form factor $\Delta_{\hat{R}}$. As
the $\overline{B}\left(\Phi_{B}\right)$ term is simply the NLO~differential
 cross section integrated over the radiative variables,
it is naturally positive, which is reflected in the absence of events
with negative weights, which are an unpleasant feature of other
next-to-leading order matching schemes. 

Since any further emissions constitute higher-order terms in the differential
cross section, next-to-next-to-leading order and beyond, we may generate
higher multiplicities in the usual way, by showering the radiative
events using the standard parton shower algorithm. These showers must
not generate emissions with $p_{T}$ greater than that of the emitted
parton in the event generated according to Eq.\,\ref{eq:powheg_5},%
\footnote{Should the \textsf{POWHEG} event turn out to be a non-radiative event, the
$p_{T}$ veto scale is zero so these events remain as no-emission
events.%
} they must be \emph{vetoed showers}. For a parton shower which evolves
in $p_{T}$ the \textsf{POWHEG} implementation is a trivial task; one simply
initiates a parton shower from each external leg of the radiative
\textsf{POWHEG} event using its $p_{T}$ as the initial-evolution scale. 

Angular-ordered parton showers account for the phenomenon of QCD coherence
where wide-angle soft gluon emissions, from near collinear configurations
of two or more partons, have insufficient transverse resolving power
to be sensitive to the constituent emitters. In effect the resulting
radiation pattern is determined by the colour charge and momentum of the
\emph{mother} of the emitters, rather than the emitters themselves.
Ordering branchings in the parton shower in terms of the angle between
the branching products takes this logarithmically enhanced effect
into account; wide-angle soft partons are emitted from the mother
of any subsequent smaller angle splittings by construction. 

As well as circumventing the problem of negative event weights, a second
major success of the \textsf{POWHEG} method is in defining how the highest
$p_{T}$ emission may
be modified to include the logarithmically enhanced effects of this
soft wide-angle radiation. In Ref.\,\cite{Nason:2004rx} it was shown
how the angular-ordered parton
shower which produces the hardest emission, may be decomposed into
a \emph{truncated} \emph{shower} simulating coherent, soft wide-angle
emissions, followed by the highest $p_{T}$ (hardest) emission, followed
again by further \emph{vetoed parton showers}, comprising of lower
$p_{T}$, smaller angle emissions. Performing this decomposition established
the form of the truncated and vetoed showers, thereby describing all
of the ingredients necessary to shower the radiative events in the \textsf{POWHEG} approach. 
This procedure
was proven in~\cite{Frixione:2007vw} to give agreement
with the NLO cross section, for all inclusive observables, while retaining
the logarithmic accuracy of the shower.

In the \textsf{POWHEG} framework positive weight events distributed with NLO
accuracy can be showered to resum further logarithmically
enhanced corrections by:
\begin{itemize}
\item generating an event according to Eq.\,\ref{eq:powheg_5};
\item directly hadronizing non-radiative events;
\item mapping the radiative variables parametrizing the emission
into the evolution scale, momentum fraction and azimuthal angle
$\left(\tilde{q}_{h},\, z_{h},\,\phi_{h}\right)$, from which the
parton shower would reconstruct identical momenta;
\item using the original leading-order configuration from 
      $\overline{B}\left(\Phi_{B}\right)$ evolve the leg emitting the extra radiation
      from the default initial
      scale down to $\tilde{q}_{h}$ using the truncated shower;
\item inserting a branching with 
      parameters~\mbox{$\left(\tilde{q}_{h},\, z_{h},\,\phi_{h}\right)$}
      into the shower when the evolution scale reaches $\tilde{q}_{h}$;
\item generating $p_{T}$ vetoed showers from all external legs.
\end{itemize}
This procedure allows the generation of the truncated shower for the first time
with only a few changes to the normal \textsf{Herwig++} shower algorithm.

\section{Next-to-Leading Order Cross Section\label{sec:Next-to-leading-order-cross}}

Although the NLO cross section for the Drell-Yan process was calculated nearly 30~years 
ago~\cite{Altarelli:1978id,KubarAndre:1978uy}, we have implemented an independent calculation of it more suited to our present goal, including the
decay of the vector boson and  $\gamma/Z$ interference effects. 
In this section we collect the ingredients that arise in the NLO calculation 
for $q+\bar{q}\rightarrow l+\bar{l}$, necessary to describe our implementation
of the \textsf{POWHEG} method.

\subsection{Kinematics and phase space\label{sub:Kinematics-and-phase}}

The leading-order process under study is of the type, $\bar{p}_{\oplus}+\bar{p}_{\ominus}\rightarrow\bar{p}_{1}+...+\bar{p}_{N}$,
in which all particles in the \emph{N}-body final state are colourless.
We denote the incoming hadron momenta $P_{\splusminus}$, for hadrons
incident in the $\pm z$ directions, respectively. The corresponding
massless parton momenta, with momentum fractions $\bar{x}_{\oplus}$
and $\bar{x}_{\ominus}$, are given by $\bar{p}_{\splusminus}=\bar{x}_{\splusminus}P_{\splusminus}$.
The momenta of the particles produced in the leading-order \emph{N}-body
process are $\bar{p}_{i}$, where $i$ ranges from $1$ to $N$.

We use $\Phi_{B}$ to denote a set of variables defining a point in
the \emph{N}-body phase space and convolution over the incoming momentum fractions,
and $\hat{\Phi}_{B}$ to denote a set
of variables parametrizing the \emph{N}-body phase space in the centre-of-mass
frame of the partonic process. It will also be convenient to define
$\bar{p}$ as the total momentum of the colour neutral particles,
$\bar{p}\equiv\bar{x}_{\oplus}P_{\oplus}+\bar{x}_{\ominus}P_{\ominus}$,
and $\bar{y}$ as the rapidity of $\bar{p}$. The partons'
momentum fractions are therefore 
\begin{align}
\bar{x}_{\oplus} & =\sqrt{\frac{\bar{p}^{2}}{s}}\mathrm{e}^{+\bar{y}}\,, & \bar{x}_{\ominus} & =\sqrt{\frac{\bar{p}^{2}}{s}}\mathrm{e}^{-\bar{y}}.\label{eq:nlo_1_1}
\end{align}
 The phase space for the leading-order process is
\begin{equation}
\mathrm{d}\Phi_{B}\mbox{ }=\mbox{ }\mathrm{d}\bar{x}_{\oplus}\,\mathrm{d}\bar{x}_{\ominus}\,\mathrm{d}\hat{\Phi}_{B}\mbox{ }=\mbox{ }\frac{1}{s}\,\mathrm{d}\bar{p}^{2}\,\mathrm{d}\bar{y}\,\mathrm{d}\hat{\Phi}_{B},\label{eq:nlo_1_2}\end{equation}
 where $\mathrm{d}\hat{\Phi}_{B}$ is the Lorentz invariant phase space for
the partonic process, in $n=4-2\epsilon$ dimensions,
\begin{align}
\mathrm{d}\hat{\Phi}_{B} & =\left(2\pi\right)^{n}\,\prod_{i}\,\left[\mathrm{d}\bar{p}_{i}\right]\,\delta^{n}\left(\bar{p}_{\oplus}+\bar{p}_{\ominus}-\bar{p}\right)\,, & \left[\mathrm{d}\bar{p}_{i}\right] & =\frac{\mathrm{d}^{n-1}\bar{p}_{i}}{\left(2\pi\right)^{n-1}2\bar{E}_{i}}\,,\label{eq:nlo_1_3}\end{align}
 and $s$ the hadronic centre-of-mass energy. The set of variables
$\Phi_{B}=\left\{ \bar{p}^{2},\,\bar{y},\,\hat{\Phi}_{B}\right\} $
defines the \emph{Born variables}.

The real emission corrections to the
leading-order process consist of~$2\rightarrow N+1$\linebreak processes, $p_{\oplus}+p_{\ominus}\rightarrow p_{1}+...+p_{N}+k$,
where we denote the momenta of the \emph{N} final-state colourless
particles $p_{i}$ and that of the extra colour charged parton by
$k$. The momentum fractions of the incoming partons are distinguished
from those in the $2\rightarrow N$ process as $x_{\oplus}$ and $x_{\ominus}$
($p_{\splusminus}=x_{\splusminus}P_{\splusminus}$). For these processes
we introduce the Mandelstam variables $\hat{s}$, $\hat{t}$, $\hat{u}$
and the related \emph{radiative variables} $\Phi_{R}=\left\{ x,\, v,\,\phi\right\} $,
which parametrize the extra emission:
\begin{subequations}
\begin{eqnarray}
\hat{s} & = & \left(p_{\oplus}+p_{\ominus}\right)^{2} =  \frac{p^{2}}{x}\,,\\
\hat{t} & = & \left(p_{\oplus}-k\right)^{2} \ \; =  \frac{p^{2}}{x}\left(x-1\right)\left(1-v\right)\,,\\
\hat{u} & = & \left(p_{\ominus}-k\right)^{2}\ \; =  \frac{p^{2}}{x}\left(x-1\right)v\,,
\end{eqnarray}\label{eq:nlo_1_4}\end{subequations}
where $\phi$ is the azimuthal angle of $k$ with respect to the $+z$
axis, and $p$ the total momentum of the colourless particles, $p\equiv x_{\oplus}P_{\oplus}+x_{\ominus}P_{\ominus}-k$.

To perform a simultaneous Monte Carlo sampling of the \emph{N}- and
\emph{N}+1-body phase spaces one has to specify the integration variables.
We choose two of these to be the mass and rapidity of the system of
colourless particles, therefore $\bar{p}^{2}\equiv p^{2}$
and $\bar{y}\equiv y$, where $y$ is defined by analogy to $\bar{y}$
as the rapidity of $p$.%
\footnote{Henceforth we will always refer to these variables as $p^{2}$ and
$y$.%
} The momentum fractions of the partons for
$2\rightarrow N+1$ processes are therefore related to those of the $2\rightarrow N$
process by
\begin{align}
x_{\oplus} & =\frac{\bar{x}_{\oplus}}{\sqrt{x}}\sqrt{\frac{1-\left(1-x\right)\left(1-v\right)}{1-\left(1-x\right)v}}\,, & x_{\ominus} & =\frac{\bar{x}_{\ominus}}{\sqrt{x}}\sqrt{\frac{1-\left(1-x\right)v}{1-\left(1-x\right)\left(1-v\right)}}.\label{eq:nlo_1_5}
\end{align}

The phase space of the \emph{N}+1-body real emission processes can be written in 
$n=4-2\epsilon$ dimensions as
\begin{equation}
\mathrm{d}\Phi_{N+1}=\mathrm{d}\Phi_{B}\,\mathrm{d}\Phi_{R}\,\frac{p^{2}}{\left(4\pi\right)^{2}x^{2}}\,\left(\frac{4\pi}{p^{2}}\right)^{\epsilon}\frac{1}{\Gamma\left(1-\epsilon\right)}\,\mbox{ }\mathcal{J}\left(x,v\right),\label{eq:nlo_1_17}
\end{equation}
 where here the partonic Born variables $\hat{\Phi}_{B}$ specify a configuration
 in the rest frame of $p$ rather than $\bar{p}$.
 The function $\mathcal{J}\left(x,v\right)$ is given by
\begin{equation}
\mathcal{J}\left(x,v\right) = \left[\mathcal{S}\delta\left(1-x\right)+\mathcal{C}\left(x\right)\left(\delta\left(v\right)+\delta\left(1-v\right)\right)+\mathcal{H}\left(x,v\right)\right]\mbox{ }v\left(1-v\right)\left(1-x\right)^{2},\label{eq:nlo_1_15}
\end{equation}
 where
\begin{subequations}\begin{eqnarray}
\mathcal{S} & = & \frac{1}{\epsilon^{2}}-\frac{\pi^{2}}{6},\\
\mathcal{C}\left(x\right) & = & -\frac{1}{\epsilon}\frac{1}{\left(1-x\right)_{+}}-\frac{1}{\left(1-x\right)_{+}}\ln x+2\left(\frac{\ln\left(1-x\right)}{1-x}\right)_{+},\\
\mathcal{H}\left(x,v\right) & = & \frac{1}{\left(1-x\right)_{+}}\left(\frac{1}{v_{+}}+\frac{1}{\left(1-v\right)_{+}}\right).
\end{eqnarray}\label{eq:nlo_1_16}\end{subequations}
 The labelling $\mathcal{S}$, $\mathcal{C}$, $\mathcal{H}$ reflects
the fact that the $\mathcal{S}$ and $\mathcal{C}$ terms are multiplied
by $\delta$-functions which limit their contributions to configurations
with \emph{soft} $\left(x\rightarrow1\right)$ and \emph{collinear}
$\left(v\rightarrow0,1\right)$ emissions, while $\mathcal{H}$ is
not associated with soft or collinear configurations but instead contributes
to \emph{hard} emissions%
\footnote{Here by hard we simply mean emissions which are neither soft or collinear.%
} of the extra parton $k$.

 The radiative phase space can be parametrized in terms of the radiative
variables
\begin{equation}
\mathrm{d}\Phi_{R} = \frac{1}{2\pi}\,\mathrm{d}x\,\mathrm{d}v\,\mathrm{d}\phi\,,
\end{equation}
 where, in the partonic centre-of-mass frame, $x=1-k^{0}/E$, with
$E$ the energy of either of the colliding partons, 
and $v=\frac{1}{2}\left(1+\cos\theta\right)$,
with $\theta$ and $\phi$ the polar and azimuthal angles of $k$
with respect to the $+z$ axis.

As the rapidity of $p$ and $\bar{p}$ are equal, it is always
possible to define a boost \mbox{$\mathbb{B}=\mathbb{B}_{L}^{-1}\mathbb{B}_{T}\,\mathbb{B}_{L}$},
such that $\mathbb{B}\mbox{ }p=\bar{p}$, where $\mathbb{B}_{L}$
is a longitudinal boost to the frame in which $y=0$ and $\mathbb{B}_{T}$
is a boost in the transverse direction, to the frame in which the
transverse momentum of $p$ is zero. It follows that an \emph{N}+1-body
configuration can be assembled by first reconstructing the \emph{N}-body
configuration corresponding to $\Phi_{B}$, then $p$ and $k$ (from
$p^{2},$ $y$, $\Phi_{R}$), at which point the boost $\mathbb{B}$
can be calculated and its inverse applied to the \emph{N}-body configuration.
Although $\mathbb{B}_{L}$ is uniquely defined due to $p$ and $\bar{p}$
having the same rapidity, $y$, the transverse boost $\mathbb{B}_{T}$
can be modified according to $\mathbb{B}_{T}\rightarrow\mathbb{R}\mathbb{B}_{T}$,
with $\mathbb{R}$ a rotation, and $\mathbb{B}$ will still satisfy
$\mathbb{B}\, p=\bar{p}$. A convention must be adopted
to fix $\mathbb{B}_{T}$, we shall return to this point later.

\subsection{Matrix elements}

The squared, spin and colour averaged leading-order matrix
element for the $q+\bar{q}\rightarrow l+\bar{l}$ Drell-Yan process is given
by $\mathcal{M}_{q\bar{q}}^{B}\left(p_{q},p_{\bar{q}}\right)$, where
the first (second) argument refers to the incoming fermion (antifermion)
momentum. The real emission radiative corrections consist of three
processes: $q+\bar{q}\rightarrow l+\bar{l}+g$; $g+q\rightarrow l+\bar{l}+q$;
and $g+\bar{q}\rightarrow l+\bar{l}+\bar{q}$. The matrix elements squared for 
these processes are 
given by\begin{subequations}\begin{eqnarray}
\mathcal{M}_{q\bar{q}}^{R} & = & \frac{\mathcal{N}_{q\bar{q}}}{p^{2}\hat{t}\hat{u}}\left[\left(\hat{s}+\hat{t}\right)^{2}\mathcal{M}_{q\bar{q}}^{B}\left(\tilde{p}_{q},\tilde{p}_{\bar{q}g}\right)+\left(\hat{s}+\hat{u}\right)^{2}\mathcal{M}_{q\bar{q}}^{B}\left(\tilde{p}_{qg},\tilde{p}_{\bar{q}}\right)\right.\\
 &  & \left.\,\,\,\,\,\,\,\,\,\,\,\,\,\,\,\,\,\,\,\,\,\,\,\,\,\,\,\,\,\,-\frac{1}{2}\epsilon\left(\hat{t}+\hat{u}\right)^{2}\left(\mathcal{M}_{q\bar{q}}^{B}\left(\tilde{p}_{g},\tilde{p}_{gg}\right)+\mathcal{M}_{q\bar{q}}^{B}\left(\tilde{p}_{gg},\tilde{p}_{g}\right)\right)\right],\nonumber \\
\mathcal{M}_{qg}^{R} & = & -\frac{\mathcal{N}_{qg}}{p^{2}\hat{u}\hat{s}}\left[\left(\hat{t}+\hat{u}\right)^{2}\mathcal{M}_{q\bar{q}}^{B}\left(\tilde{p}_{qg},\tilde{p}_{\bar{q}}\right)+\left(\hat{t}+\hat{s}\right)^{2}\mathcal{M}_{q\bar{q}}^{B}\left(\tilde{p}_{q},\tilde{p}_{\bar{q}g}\right)\right.\\
 &  & \left.\,\,\,\,\,\,\,\,\,\,\,\,\,\,\,\,\,\,\,\,\,\,\,\,\,\,\,\,\,\,-\frac{1}{2}\epsilon\left(\hat{u}+\hat{s}\right)^{2}\left(\mathcal{M}_{q\bar{q}}^{B}\left(\tilde{p}_{g},\tilde{p}_{gg}\right)+\mathcal{M}_{q\bar{q}}^{B}\left(\tilde{p}_{gg},\tilde{p}_{g}\right)\right)\right],\nonumber \\
\mathcal{M}_{g\bar{q}}^{R} & = & -\frac{\mathcal{N}_{qg}}{p^{2}\hat{s}\hat{t}}\left[\left(\hat{u}+\hat{s}\right)^{2}\mathcal{M}_{q\bar{q}}^{B}\left(\tilde{p}_{qg},\tilde{p}_{\bar{q}}\right)+\left(\hat{u}+\hat{t}\right)^{2}\mathcal{M}_{q\bar{q}}^{B}\left(\tilde{p}_{q},\tilde{p}_{\bar{q}g}\right)\right.\\
 &  & \left.\,\,\,\,\,\,\,\,\,\,\,\,\,\,\,\,\,\,\,\,\,\,\,\,\,\,\,\,\,\,-\frac{1}{2}\epsilon\left(\hat{s}+\hat{t}\right)^{2}\left(\mathcal{M}_{q\bar{q}}^{B}\left(\tilde{p}_{g},\tilde{p}_{gg}\right)+\mathcal{M}_{q\bar{q}}^{B}\left(\tilde{p}_{gg},\tilde{p}_{g}\right)\right)\right],\nonumber \end{eqnarray}\label{eq:nlo_2_1}\end{subequations}
where, for a more uniform notation, and to help show how the crossing of the leading-order
process is manifest, we have denoted the final-state quark momentum
in the $qg$ initiated process by $p_{\bar{q}}$ and the final-state
antiquark momentum in the $g\bar{q}$ process by $p_{q}$. The shifted
momenta $\tilde{p}_{i}$, $\tilde{p}_{jg}$, and the normalization
constants $\mathcal{N}_{q\bar{q}}$, $\mathcal{N}_{qg}$ are given
by\begin{align}
\tilde{p}_{i} & =\frac{1}{x_{i}}p_{i}\,, & \mathcal{N}_{q\bar{q}} & =8\pi\alpha_{S}C_{F}\mu^{2\epsilon}\,,\nonumber \\
x_{i} & =\frac{2p.p_{i}}{p^{2}}\,, & \mathcal{N}_{qg} & =8\pi\alpha_{S}T_{F}\mu^{2\epsilon}/\left(1-\epsilon\right)\,,\label{eq:nlo_2_4}\\
\tilde{p}_{jg} & =p-\tilde{p}_{i}\,,\nonumber \end{align}
 where $\mu$ is the regularization  scale emerging from the use of
conventional dimensional regularization. The shifted momenta satisfy 
\begin{align}
\tilde{p}_{i}^{2} & =\tilde{p}_{jg}^{2}=0, & \tilde{p}_{i}+\tilde{p}_{jg} & =p,
\label{eq:nlo_2_5}
\end{align}
\emph{i.e.} they obey the relations required for them to be considered
as describing a kinematic configuration for the leading-order process,
hence they form valid arguments for $\mathcal{M}_{q\bar{q}}^{B}$.

These matrix elements hold independently of the type of exchanged
vector boson, they also hold for the case that the leading-order process
consists of interferences between diagrams with different exchanged
vector bosons. This universal behaviour is due to the factorisation
of the NLO hadron tensor, into kinematic factors multiplying the 
leading-order hadron tensor. Such a factorisation of the matrix element is
not necessary for the implementation of the \textsf{POWHEG} method but it improves
the flexibility and generality of our implementation of the Drell-Yan process.
This allows us generate \emph{N}+1-body configurations, according
to the full real emission matrix element, given the leading-order
configuration using a technique of sampling the radiative phase space with a branching
algorithm known as \emph{the Kleiss trick}~\cite{Kleiss:1986re,Seymour:1994we},
which we have extended to $n$ dimensions as needed for a complete NLO
calculation, as we now describe in detail.

The $\mathcal{O}\left(\alpha_{S}\right)$ virtual corrections to the
$q+\bar{q}\rightarrow l+\bar{l}$ process solely consist of
the $q\bar{q}$ vertex correction. At NLO, this loop diagram only
contributes to the matrix element through its interference with the
leading-order amplitude, correcting it by,
\begin{equation}
\mathcal{M}_{q\bar{q}}^{V}=\frac{\alpha_{S}C_{F}}{2\pi}\left(\frac{4\pi\mu^{2}}{p^{2}}\right)^{\epsilon}\frac{1}{\Gamma\left(1-\epsilon\right)}\left[-\frac{2}{\epsilon^{2}}-\frac{3}{\epsilon}-8+\pi^{2}\right]\mathcal{M}_{q\bar{q}}^{B}\left(p_{q},p_{\bar{q}}\right).\label{eq:nlo_2_6}\end{equation}

\subsection{Differential cross section}

The partonic flux due to parton $a$ in hadron $A$ and
parton $b$ in hadron $B$, at scale $\mu^{2}$, with momentum fractions
$x_{\oplus}$ and $x_{\ominus}$ respectively, is defined as
\begin{equation}
\mathcal{L}_{ab}\left(x_{\oplus},x_{\ominus}\right)=f_{a}^{A}\left(x_{\oplus},\mu^{2}\right)f_{b}^{B}\left(x_{\ominus},\mu^{2}\right).\label{eq:nlo_3_1}\end{equation}
 In $\mathcal{L}_{ab}\left(x_{\oplus},x_{\ominus}\right)$ the functions
$f_{i}^{I}\left(x_{i},\mu^{2}\right)$ are the parton distribution
functions~(PDFs) for finding a parton $i$ in hadron $I$ with momentum
fraction $x_{i}$ at scale $\mu^{2}$. The contribution to the differential
cross section from the leading-order process $q+\bar{q}\rightarrow l+\bar{l}$
is therefore
\begin{equation}
\mathrm{d}\sigma_{q\bar{q}}^{B} = B\left(\Phi_{B}\right)\,\mathrm{d}\Phi_{B}\,,
\end{equation}
where
\begin{equation}
B\left(\Phi_{B}\right) = \frac{1}{2p^{2}}\,\mathcal{M}_{q\bar{q}}^{B}\left(\bar{p}_{\oplus},\bar{p}_{\ominus}\right)\,\mathcal{L}_{q\bar{q}}\left(\bar{x}_{\oplus},\bar{x}_{\ominus}\right).\label{eq:nlo_3_2}
\end{equation}
 The virtual corrections (Eq.\,\ref{eq:nlo_2_6}) add
\begin{equation}
\mathrm{d}\sigma_{q\bar{q}}^{V} = V_{0}\left(\Phi_{B}\right)\,\mathrm{d}\Phi_{B}\,,\label{eq:nlo_3_3}
\end{equation}
where
\begin{equation} V_{0}\left(\Phi_{B}\right) = \frac{\alpha_{S}C_{F}}{2\pi}\left[-\frac{2}{\epsilon}\left(\frac{1}{\bar{\epsilon}}+\ln\left(\frac{\mu^{2}}{p^{2}}\right)\right)-\frac{3}{\bar{\epsilon}}+\frac{\pi^{2}}{3}+\mathcal{V}_{q\bar{q}}\right]\, B\left(\Phi_{B}\right)\,,
\end{equation}
 with $\bar{\epsilon}$ defined by $\frac{1}{\bar{\epsilon}}=\frac{1}{\epsilon}-\gamma_{E}+\ln\left(4\pi\right)$ and
\begin{equation}
\mathcal{V}_{q\bar{q}}=-3\ln\left(\frac{\mu^{2}}{p^{2}}\right)+\frac{2\pi^{2}}{3}-8\,.
\end{equation}
In Eqs.\,\ref{eq:nlo_3_2} and \ref{eq:nlo_3_3} we have included a
flux factor $1/2p^{2}$ for the partonic process. The subscript on
$V_{0}$ identifies the bare divergent quantity.

The differential cross section for the real emission processes 
$a+b\rightarrow l+\bar{l}+c$ is of the form
\begin{equation}
\mathrm{d}\sigma_{ab}^{R} =\frac{1}{4\pi x}\mbox{ }\mathcal{L}_{ab}\left(x_{\oplus},x_{\ominus}\right)\mbox{ }\left[\sum_{i}\mbox{ }f_{i}\left(x,v\right)\mbox{ }\mathcal{M}_{q\bar{q}}^{B}\left(\tilde{p}_{i},\tilde{p}_{j}\right)\right]\mbox{ }\mathrm{d}\Phi_{B}\mbox{ }\mathrm{d}\tilde{\Phi}_{R}\,,\label{eq:nlo_3_4}
\end{equation}
 where for $i=qg,$ $q$, $g$, $gg$, we have $j\left(i\right)=\bar{q}$,
$\bar{q}g$, $gg$, $g$. Each of the $f_{i}\left(x,v\right)$ functions
is defined as the coefficient of the squared leading-order matrix
element in the $\mathcal{M}_{ab}^{R}$ real-emission matrix element,
Eq.\,\ref{eq:nlo_2_1}.

Returning to the earlier discussion of the definition of the Born
variables and the boost $\mathbb{B}$, we note that for each $i$
in Eq.\,\ref{eq:nlo_3_4} the boost to the rest frame of $\tilde{p}_{i}+\tilde{p}_{j}$
is the same up to an overall rotation (as $\tilde{p}_{i}+\tilde{p}_{j}=p$).
Furthermore, for $i=q$ we have $\tilde{p}_{i}=p_{i}/x_{i}$, so $\tilde{p}_{i}$
and $p_{i}$ have the same orientation, as do $\tilde{p}_{j}$ and
$p_{j}$ for index $j=\bar{q}$. Since $\mathbb{B}$ was defined up
to an arbitrary rotation $\mathbb{R}$, we could choose to resolve
this ambiguity by setting $\mathbb{R}=\mathbb{R}_{q}$, $\mathbb{B}=\mathbb{B}_{q}$,
where $\mathbb{B}_{q}$ additionally satisfies,
\begin{equation}
\mathbb{B}_{q}\,\tilde{p}_{q}\equiv\mathbb{B}_{L}^{-1}\mathbb{R}_{q}\,\mathbb{B}_{T}\,\mathbb{B}_{L}\,\tilde{p}_{q}=\bar{p}_{\oplus},\label{eq:nlo_3_5}
\end{equation}
\emph{i.e.} $\mathbb{R}_{q}$ is such that the value of $\tilde{p}_{q}$
in the $p$ rest frame is equal to $\bar{p}_{\oplus}$ in the $\bar{p}$
rest frame, therefore
\begin{equation}
\mathcal{M}_{q\bar{q}}^{B}\left(\tilde{p}_{q},\tilde{p}_{\bar{q}g}\right)\mathrm{d}\Phi_{B}\equiv\mathcal{M}_{q\bar{q}}^{B}\left(\bar{p}_{\oplus},\bar{p}_{\ominus}\right)\mathrm{d}\Phi_{B}.\label{eq:nlo_3_6}\end{equation}
 To have the analogous equivalence for the $i,\, j=qg,\,\bar{q}$
term, one needs to have $\mathbb{B}=\mathbb{B}_{qg}$, defined such
that $\mathbb{B}_{qg}\,\tilde{p}_{\bar{q}}=\bar{p}_{\ominus}$ is
satisfied along with original requirement ($\mathbb{B}\, p=\bar{p}$).

Similar boosts can be constructed for the $i=g,\, gg$
terms in the matrix element, mapping $\tilde{p}_{g}$ to $\bar{p}_{\oplus}$
and $\bar{p}_{\ominus}$ respectively. However, as these terms carry
a factor $\epsilon$, they only contribute to exactly collinear configurations
in which the parton $k$ is \emph{unresolved}, hence those boosts
are a purely formal consideration, and are not needed in practice.

The real emission phase space can therefore be sampled
using a simple Monte Carlo \emph{branching algorithm}; namely, given
a set of Born and radiative variables, the event is reconstructed
as described in Sect.\,\ref{sub:Kinematics-and-phase}, except that,
where previously a single boost $\mathbb{B}^{-1}$ was used to embed
the \emph{N}-body configuration in the \emph{N}+1-body event, now
we use a boost $\mathbb{B}_{i}^{-1}$ selected according to the probability
distribution $\mathcal{P}_{i}=f_{i}\left(x,v\right)/\sum_{i}\, f_{i}\left(x,v\right)$.
Sampling the phase space in this way, the generation of the Born variables
becomes completely independent of the radiative variables:
\begin{subequations}\begin{eqnarray}
\mathrm{d}\sigma_{ab}^{R} & = & R_{ab,0}\left(\Phi_{B},\Phi_{R}\right)\,\mathrm{d}\Phi_{R}\,\mathrm{d}\Phi_{B}\,\mbox{ }\\
R_{ab,0}\left(\Phi_{B},\Phi_{R}\right) & = & \frac{\hat{s}}{2\pi}\,\hat{\mathcal{L}}_{ab}\left(x_{\oplus},x_{\ominus}\right)\,\sum_{i}\mbox{ }f_{i}\left(x,v\right)\,\frac{\mathrm{d}\tilde{\Phi}_{R}}{\mathrm{d}\Phi_{R}}\, B\left(\Phi_{B}\right),
\end{eqnarray}\label{eq:nlo_3_7}\end{subequations}
 where $\hat{\mathcal{L}}_{ab}$ is the ratio of the general parton
flux relative to that of the leading-order process,
\begin{equation}
\hat{\mathcal{L}}_{ab}\left(x_{\oplus},x_{\ominus}\right)=\frac{\mathcal{L}_{ab}\left(x_{\oplus},x_{\oplus}\right)}{\mathcal{L}_{q\bar{q}}\left(\bar{x}_{\oplus},\bar{x}_{\ominus}\right)}.\label{eq:nlo_3_8}
\end{equation}
The functions $R_{ab,0}\left(\Phi_{B},\Phi_{R}\right)$, calculated
according to Eq.\,\ref{eq:nlo_3_7}, are given by
\begin{equation}
R_{ab,0}\left(\Phi_{B},\Phi_{R}\right) = \frac{\alpha_{S}C_{ab}}{2\pi}\,\frac{1}{x}\,\hat{\mathcal{L}}_{ab}\left(x_{\oplus},x_{\ominus}\right)\,\mathcal{R}_{ab,0}\, B\left(\Phi_{B}\right),
\label{eq:nlo_3_9}
\end{equation}
 where $C_{ab}=C_{F}$ for $ab=q\bar{q}$ and $T_{F}$ otherwise.

For the $q\bar{q}$ contribution $\mathcal{R}_{q\bar{q},0}$ is given by
\begin{equation}
\mathcal{R}_{q\bar{q},0} = \mathcal{S}_{q\bar{q}}\delta\left(1-x\right)+\left(-\frac{1}{\bar{\epsilon}}P_{qq}+\mathcal{C}_{q\bar{q}}\right)\left(\delta\left(v\right)+\delta\left(1-v\right)\right)+\mathcal{H}_{q\bar{q}},
\end{equation}
where
\begin{subequations}\begin{eqnarray}
\mathcal{S}_{q\bar{q}} & = & \left(\frac{2}{\epsilon}+3\right)\left(\frac{1}{\bar{\epsilon}}+\ln\left(\frac{\mu^{2}}{p^{2}}\right)\right)-\frac{\pi^{2}}{3}-3\ln\left(\frac{\mu^{2}}{p^{2}}\right)\,, \\
P_{qq} & = & \left(\frac{1+x^{2}}{1-x}\right)_{+}\,, \\
\mathcal{C}_{q\bar{q}} & = & \left(1+x^{2}\right)\left(\frac{1}{\left(1-x\right)_{+}}\ln\left(\frac{p^{2}}{\mu^{2}x}\right)+2\left(\frac{\ln\left(1-x\right)}{1-x}\right)_{+}\right)+1-x\,, \\
\mathcal{H}_{q\bar{q}} & = & \frac{1}{\left(1-x\right)_{+}}\left(\frac{1}{v_{+}}+\frac{1}{\left(1-v\right)_{+}}\right)\left(\left(1-x\right)^{2}\left(1-2v\left(1-v\right)\right)+2x\right). 
\end{eqnarray}\end{subequations}
 For the $qg$ contribution $\mathcal{R}_{qg,0}$ is given by
\begin{equation}\mathcal{R}_{qg,0} = \left(-\frac{1}{\bar{\epsilon}}\, P_{gq}+\mathcal{C}_{qg}\right)\delta\left(v\right)+\mathcal{H}_{qg}\,,\label{eq:nlo_3_11}
\end{equation}
where
\begin{subequations}\begin{eqnarray}
P_{gq} & = & x^{2}+\left(1-x\right)^{2}\,,\\
\mathcal{C}_{qg} & = & \left(x^{2}+\left(1-x\right)^{2}\right)\left(\ln\left(\frac{p^{2}}{\mu^{2}x}\right)+2\ln\left(1-x\right)\right)+2x\left(1-x\right)\,, \\
\mathcal{H}_{qg} & = & \frac{1}{v_{+}}\left(2x\left(1-x\right)v+\left(1-x\right)^{2}v^{2}+x^{2}+\left(1-x\right)^{2}\right).
\end{eqnarray}\end{subequations}
 The function $\mathcal{R}_{g\bar{q},0}$ is equal to $\mathcal{R}_{qg,0}$ under
the replacement $v\leftrightarrow1-v$.

The soft term in the real correction $R_{q\bar{q},0}$, proportional
to $\delta\left(1-x\right)$, is now combined with the virtual correction
$V_{0}$, to give a soft-virtual contribution $V$, which is 
finite for $\epsilon\rightarrow0$.
The remaining divergences are initial-state collinear divergences
proportional to $\delta\left(v\right)$ and/or $\delta\left(1-v\right)$.
Working in the $\overline{\mathrm{MS}}$ scheme these are 
absorbed into the definition of the PDFs. This renormalization and cancellation
of divergences amounts to the replacements, $V_{0}\rightarrow V$
and $R_{ab,0}\rightarrow R_{ab}$, where\begin{equation}
V\left(\Phi_{B}\right)=\frac{\alpha_{S}C_{F}}{2\pi}\,\mathcal{V}_{q\bar{q}}\, B\left(\Phi_{B}\right)\label{eq:nlo_3_12}\end{equation}
and
\begin{subequations}\begin{eqnarray}
R_{ab}\left(\Phi_{B},\Phi_{R}\right)\, & = & \frac{\alpha_{S}C_{ab}}{2\pi}\,\frac{1}{x}\,\mathcal{R}_{ab}\,\hat{\mathcal{L}}_{ab}\left(x_{\oplus},x_{\ominus}\right)\, B\left(\Phi_{B}\right),\label{eq:nlo_3_13}\\
\mathcal{R}_{q\bar{q}}\left(\Phi_{B},\Phi_{R}\right) & = & \mathcal{C}_{q\bar{q}}\left(\delta\left(v\right)+\delta\left(1-v\right)\right)+\mathcal{H}_{q\bar{q}}, \\
\mathcal{R}_{qg}\left(\Phi_{B},\Phi_{R}\right) & = & \mathcal{C}_{qg}\delta\left(v\right)+\mathcal{H}_{qg}, \\
\mathcal{R}_{g\bar{q}}\left(\Phi_{B},\Phi_{R}\right) & = & \mathcal{C}_{g\bar{q}}\delta\left(1-v\right)+\mathcal{H}_{g\bar{q}}. 
\end{eqnarray}\end{subequations}
 Adding these contributions we obtain the NLO differential cross section:
\begin{equation}
\mathrm{d}\sigma=B\left(\Phi_{B}\right)\mathrm{d}\Phi_{B}+V\left(\Phi_{B}\right)\mathrm{d}\Phi_{B}+R\left(\Phi_{B},\Phi_{R}\right)\mathrm{d}\Phi_{B}\mathrm{d}\Phi_{R}\,,\label{eq:nlo_3_14}\end{equation}
 where $R$ denotes the sum of the $R_{ab}$ terms.

\section{Implementation in \HWPP\ \label{sec:Implementation}}

In the first two parts of this section we describe how distributions
of NLO accurate non-radiative and single emission events are generated.
In Sect.~\ref{sub:Truncated-and-vetoed}
we describe the simulation of further, lower $p_{T}$, emissions,
from the radiative events, using the truncated and vetoed
shower algorithms.

\subsection{Generation of the leading-order configuration\label{sub:genlo}}

As noted in Sect.~\ref{sec:The-POWHEG-method}, sampling the $\overline{B}\left(\Phi_{B}\right)$
function (Eq.\,\ref{eq:powheg_6}), which is the next-to-leading
order differential cross section integrated over the radiative variables,
\begin{equation}
\overline{B}\left(\Phi_{B}\right) = B\left(\Phi_{B}\right)\left[1+\frac{\alpha_{S}C_{F}}{2\pi}\,\mathcal{V}_{q\bar{q}}+\sum_{ab}\int\mathrm{d}\Phi_{R}\,\frac{\alpha_{S}C_{ab}}{2\pi}\,\frac{1}{x}\,\mathcal{R}_{ab}\,\hat{\mathcal{L}}_{ab}\left(x_{\oplus},x_{\ominus}\right)\right]\label{eq:bbar_impl_1}
\end{equation}
provides Born variables
$\Phi_{B}$ distributed according to the exact NLO differential
cross section. The way in which the leading-order process is
factorised inside the real emission terms $R_{ab}$ allows
the $\overline{B}\left(\Phi_{B}\right)$
distribution to be generated as a straightforward reweighting%
\footnote{Apart from the requirement that the final state be colour neutral,
this reweighting is independent of the details of the vector
boson and any decay it undergoes.%
} of the leading-order cross section.

For convenience the radiative phase space $\mathrm{d}\Phi_{R}$ is
reparametrized by variables on the interval $\left[0,1\right]$ such
that the radiative phase space volume is unity, a three-dimensional
unit cube. This is achieved by a trivial change of variables $\phi\rightarrow\bar{\phi}=\phi/2\pi$
and $x\rightarrow\tilde{x}$, where $\tilde{x}$ is defined by \begin{equation}
x\left(\tilde{x},v\right)=\bar{x}\left(v\right)+\left(1-\bar{x}\left(v\right)\right)\tilde{x},\label{eq:bbar_impl_2}\end{equation}
where $\bar{x}(v)$ is the lower limit on the $x$ integration.
Numerical implementation of the $\overline{B}\left(\Phi_{B}\right)$
distribution requires all plus distributions be replaced by regular
functions, the results of which are given in Appendix.\,\ref{sec:Plus-distributions}. 

 The generation of the leading-order, \emph{N}-body configuration
proceeds as follows: 
\begin{enumerate}
\item a leading-order configuration is generated using the standard
\HWPP\ leading-order matrix element generator, providing the Born variables
$\Phi_{B}$ with an associated weight $B\left(\Phi_{B}\right)$;
\item radiative variables $\Phi_{R}$ are then generated by sampling $\overline{B}\left(\Phi_{B}\right)$,
parametrized in terms of the `unit-cube' variables $\tilde{x}$, $v$,
$\bar{\phi}$, using the Auto-Compensating Divide-and-Conquer
(ACDC) phase space generator \cite{Lonnblad:2006pt}, which implements
a variant of the VEGAS algorithm \cite{Lepage:1980dq}; 
\item the leading-order configuration is accepted with a probability proportional
to the integrand of Eq.\,\ref{eq:bbar_impl_1} evaluated at $\left\{ p^{2},\, y,\,\Phi_{R}\right\} $. 
\end{enumerate}

\subsection{Generation of the hardest emission\label{sub:genhard}}

The hardest (highest $p_{T}$) emission is generated from the \emph{N}-body
configuration according to the modified Sudakov form factor, Eq.\,\ref{eq:powheg_4}.
The integrand in the exponent of the Sudakov form factor consists of
three different contributions, one for each channel $ab$=$q\bar{q}$,
$qg$, $g\bar{q}$. The integrands are defined as 
\begin{equation}
W_{ab}\left(\Phi_{R},\Phi_{B}\right)=\frac{\hat{R}_{ab}\left(\Phi_{B},\Phi_{R}\right)}{B\left(\Phi_{B}\right)}=\frac{\alpha_{S}C_{ab}}{2\pi}\,\frac{1}{x}\,\hat{\mathcal{H}}_{ab}\,\hat{\mathcal{L}}_{ab}\left(x_{\oplus},x_{\ominus}\right),\label{eq:hardest_1}\end{equation} 
where $\hat{\mathcal{H}}_{ab}$ is equal to $\mathcal{H}_{ab}$ without
the plus prescription.
 We use the following parametrization of the partonic, vector
boson and jet momentum in the hadronic centre-of-mass frame: 
\begin{align}
p_{\oplus} & =\frac{1}{2}\sqrt{s}\left(x_{\oplus},0,0,+x_{\oplus}\right)\,, & p & =\left(m_{T}\cosh y,\, \phantom{-}p_{T}\sin\phi,\, \phantom{-}p_{T}\cos\phi,\, m_{T}\sinh y\right),\nonumber\\
p_{\ominus} & =\frac{1}{2}\sqrt{s}\left(x_{\ominus},0,0,-x_{\ominus}\right)\,, & k & =\left(p_{T}\cosh y_{k},\, -p_{T}\sin\phi,\, -p_{T}\cos\phi,\, p_{T}\sinh y_{k}\right),\end{align}
 where $p_{T}$ is the transverse momentum, $m_{T}=\sqrt{p^{2}+p_{T}^{2}}$
and $y_{k}$ is the rapidity of the additional parton. Instead of
generating the hardest emission in terms of $\Phi_{R}=\left\{ x,\, v,\,\phi\right\} $
we find it more convenient to make a change of variables to $\Phi_{R}^{\prime}=\left\{ p_{T},\, y_{k},\,\phi\right\} $,
related to $\Phi_{R}$ according to \begin{align}
p_{T}^{2} & =\frac{p^{2}}{x}\, v\left(1-v\right)\left(1-x\right)^{2}\,, & y_{k} & =y+\frac{1}{2}\ln\left(\frac{v}{1-v}\right)-\frac{1}{2}\ln\left(\frac{1-v\left(1-x\right)}{x+v\left(1-x\right)}\right).\end{align}

The modified Sudakov form factor of Eq.\,\ref{eq:powheg_4} contains
a $\theta$-function in $p_{T}$, however by choosing to parametrize
the radiative phase space in $p_{T}$ the $\theta$-function simply
becomes the upper limit on the integral. The modified Sudakov form
factor for each channel 
therefore has the form \begin{equation}
\Delta_{\hat{R}_{ab}}\left(p_{T}\right)=\exp\left(-\int_{p_{T}}^{p_{T{\rm max}}}\mathrm{d}\Phi_{R}\mbox{ }W_{ab}\left(\Phi_{R},\Phi_{B}\right)\right),\end{equation}
where $p_{T{\rm max}}$ is the maximum possible transverse momentum. The full
Sudakov form factor, $\Delta_{\hat{R}}(p_T)$, is given by the product of 
$\Delta_{\hat{R}_{ab}}(p_T)$ for the individual channels.
The radiative variables $\left(p_{T},\, y_{k}\right)$ are generated
according to Eq.\,\ref{eq:powheg_4} using the \emph{veto algorithm}.\footnote{A good description of this technique can be found in Ref.\,\cite{Sjostrand:2006za}.}
This procedure requires simple
bounding functions for each channel. Functions of the form,
\begin{equation}
g_{ab}\left(p_{T}\right)=\frac{K_{ab}}{p_{T}^{2}}\,,\end{equation}
are used, with suitable values of $K_{ab}$ for each channel together
with an overestimate of the limits for the rapidity integral, $y_{k_{{\rm min}}}$ and
$y_{k_{{\rm max}}}$.
The generation procedure then proceeds as follows:

\begin{enumerate}
\item $p_{T}$ is set to $p_{T{\rm max}}$; 
\item \label{restart} a new $\left(p_{T},\, y_{k}\right)$ configuration
is generated using two random numbers according to%
\footnote{$\mathcal{R}$ refers to a random number in the interval $[0,1]$,
a different random number is generated each time. %
} \begin{subequations} \begin{eqnarray}
p_{T} & = & {\left( \frac{1}{p_{T}}-\frac{1}{K_{ab}\left(y_{k_{{\rm max}}}-y_{k_{{\rm min}}}\right)}
\ln{\mathcal{R}} \right) } ^{-1},\\
y_{k} & = & y_{k_{{\rm min}}}+\mathcal{R}\left(y_{k_{{\rm max}}}-y_{k_{{\rm min}}}\right);\end{eqnarray}
 \end{subequations} 
\item if $p_{T}<p_{T{\rm min}}$, where $p_{T{\rm min}}$ is the minimum allowed $p_T$ for the emission, then no radiation is generated; 
\item if the generated configuration is outside of the exact phase space
boundaries then return to step \ref{restart}; 
\item if $W_{ab}\left(\Phi_{B},\Phi_{R}\right)/g_{ab}\left(p_{T}\right)>\mathcal{R}$
then accept the configuration, otherwise return to step \ref{restart}. 
\end{enumerate}
For this process there are three partonic channels contributing to
the radiative corrections, this is dealt with by using \emph{competition},
where a $\left(p_{T},\, y_{k}\right)$ configuration is generated,
as outlined above, for each channel individually and the configuration
with the highest $p_{T}$ accepted.

We employ a simple prescription \cite{Seymour:1994we}
to generate the azimuthal angle that allows the leptonic correlations
to be correctly generated. For the $q\bar{q}$ channel, the prescription
proceeds as follows:
\begin{enumerate}
\item momenta are first constructed in the vector boson rest frame; 
\item the $p_{\oplus}$ direction is chosen with probability \begin{equation}
\left(\hat{s}+\hat{t}\right)^{2}/\left(\left(\hat{s}+\hat{t}\right)^{2}+\left(\hat{s}+\hat{u}\right)^{2}\right),\label{directionChoice}\end{equation}
 otherwise the $p_{\ominus}$ direction is chosen. The momenta are
then rotated around the chosen direction by a random angle generated
uniformly on the interval $\left[0,2\pi\right]$; 
\item momenta are boosted back to the lab frame such that the rapidity of
the vector boson is the same as for the \emph{N}-body configuration. 
\end{enumerate}
The same procedure is used for the $qg$ and $g\bar{q}$ initiated channels with the replacements $\hat{s}\rightarrow\hat{t}$, $\hat{t}\rightarrow\hat{u}$, $\hat{u}\rightarrow\hat{s}$ and $\hat{s}\rightarrow\hat{u}$, $\hat{t}\rightarrow\hat{s}$, $\hat{u}\rightarrow\hat{t}$, respectively.

\subsection{Truncated and vetoed parton showers\label{sub:Truncated-and-vetoed}}

Before describing how the radiative events are further
showered, we need to recall some details of the \HWPP\
parton shower algorithm. This is described in more detail in 
Refs.\,\cite{Bahr:2008pv,Gieseke:2003rz}.
The shower starts at a scale given by the colour structure of the
hard scattering process and evolves down in the evolution variable
$\tilde{q}$ by the emission of partons in $1\rightarrow2$ branching
processes. Finally, the set of scales, $\tilde{q}$, momentum fractions,
$z$, and azimuthal angles, $\phi$, which describe these branchings,
are used to construct the momenta of all the partons radiated in the
parton shower. The \HWPP\ approach generally requires some reshuffling
of these momenta after the generation of the parton showers in order
to ensure global energy-momentum conservation.

The \emph{N}+1-body states generated as described in Sect.\,\ref{sub:genhard}
are first interpreted as a single standard
\HWPP\ shower emission, from the \emph{N}-body configuration, described
by the shower branching variables $\left(\tilde{q}_{h},\, z_{h},\,\phi_{h}\right)$.
The complete \textsf{POWHEG} shower is then performed as a single \HWPP\
shower modified by certain conditions which allows a simple but complete
implementation of the truncated shower.

 The shower algorithm proceeds as follows:
\begin{enumerate}
\item the truncated shower evolves from the hard scale, determined by the colour structure of the $N$-body process, to the hardest emission
scale $\tilde{q}_{h}$ such that the $p_{T}$ is less than that of
the hardest emission, the radiation is angular-ordered and branchings
do not change the flavour of the emitting parton; 
\item the hardest emission is forced with shower variables $\left(\tilde{q}_{h},\, z_{h},\,\phi_{h}\right)$; 
\item the shower is allowed to evolve down to the hadronization scale with
the addition of a transverse momentum veto on radiation above $p_{T_{h}}$. 
\end{enumerate}
The implementation described above requires that the hardest emission,
generated as described in Sect.\,\ref{sub:genhard}, be interpreted as
a \HWPP\ shower emission that is forced when we evolve to the
associated scale in the parton shower. In order to do this, we need
to find a mapping from the $N+1$ momenta, describing the hardest
emission, to the shower variables $\left(\tilde{q}_{h},\, z_{h},\,\phi_{h}\right)$
and an $N$-body configuration. This equates to undoing the momentum
reconstruction procedure used in the \HWPP\ shower. The
reconstruction procedure consists of two steps~\cite{Bahr:2008pv}.
First, the momenta of partons at each step of the shower are constructed,
in the centre-of-mass frame of the hadronic collision,
recursively from the shower variables.
Second, boosts are applied to each jet individually which ensures
global momentum conservation. The reconstruction process is different
for initial- and final-state radiation, here we will only consider
the initial-state case which is relevant for the Drell Yan process.

In \HWPP\ the momenta of the partons, $q_{i}$, in a jet are given
by 
\begin{equation}
q_{i}=\alpha_{i}p+\beta_{i}n+q_{\perp i},\label{eq:impl_1}
\end{equation}
where for initial-state radiation the reference vectors $p$ and
$n$ are given by the hadronic momenta of the beam particles $p_{\oplus}$
and $p_{\ominus}$ and $q_{\perp i}$ is the transverse momentum
with respect to the beam axis.

For initial-state radiation we use a backward evolution algorithm
which starts from the hard process and evolves to lower evolution scales
backwards towards the incoming hadron by the emission of time-like
partons. The reconstruction of the initial-state jet starts from the
last initial-state parton produced by the backward evolution algorithm
with momentum calculated from the fraction of the beam momentum it carries.
The momentum of the time-like daughter of this parton is reconstructed
as described in Ref.\,\cite{Bahr:2008pv}. The momentum of the space-like
daughter is then given by momentum conservation.
This process is iterated
for each initial-state branching eventually giving the momentum of
the space-like progenitor parton, colliding in the hard process.

The momenta of the two progenitor partons are then reshuffled such
that mass and rapidity of the partonic centre-of-mass system is conserved.
Under this reshuffling the progenitor momenta are transformed according
to \begin{equation}
q_{\splusminus}\rightarrow q_{\splusminus}^{\prime}=\alpha_{\splusminus}k_{\splusminus}p_{\splusminus}+\frac{\beta_{\splusminus}}{k_{\splusminus}}p_{\sminusplus}+q_{\perp\splusminus}.\label{eq:impl_2}\end{equation}
 The reshuffling parameters, $k_{\oplus}$ and $k_{\ominus}$,
are found by solving equations requiring conservation of mass and
rapidity and hence the associated Lorentz transform is obtained.

The basis vectors are the hadronic beam momenta and the $\alpha_{\splusminus}$
parameters in Eq.\,\ref{eq:impl_2} are simply the Born partonic
momentum fractions, given in Eq.\,\ref{eq:nlo_1_2}. The reshuffling
parameters, $k_{\splusminus}$, can therefore be calculated from the
momenta of the shuffled progenitors, $q_{0\splusminus}^{\prime}$.
Decomposing these momenta into their Sudakov parameters the momentum
shuffling parameters are simply 
\begin{align}
k_{\oplus} & =\frac{\alpha_{0\oplus}^{\prime}}{\bar{x}_{\oplus}}, & k_{\ominus} & =\frac{\alpha_{0\ominus}^{\prime}}{\bar{x}_{\ominus}},\label{eq:impl_3}
\end{align}
 where $\alpha^{\prime}_{0\splusminus}$ refers to the $\alpha$ parameters
in the Sudakov decomposition of the shuffled progenitors. The inverse
of the Lorentz boosts implementing the reshuffling can then be calculated
and applied to each momentum, yielding the unshuffled momenta $q_{i}$.
These are then decomposed yielding their Sudakov parameters, the
shower variables parametrizing the branching can then be determined.
The momentum fraction is given by \begin{equation}
z=\frac{\alpha_{i}}{\alpha_{\widetilde{ij}}},\label{eq:impl_4}\end{equation}
where $\alpha_{i}$ is the Sudakov parameter for the space-like parton
entering the hard process and $\alpha_{\widetilde{ij}}$ the Sudakov
parameter of the initial-state parent parton. In this simple case
the transverse momentum is simply equal to that of the off-shell space-like
parton initiating the leading-order hard process, or equivalently,
its outgoing, time-like, sister parton. The scale of the branching
is defined in terms of the $p_{T}$ and light-cone momentum fraction
$z$, as \begin{equation}
\tilde{q}^{2}=\frac{zQ_{g}^{2}+p_{T}^{2}}{\left(1-z\right)^{2}},\label{eq:impl_5}\end{equation}
 where $Q_{g}$ is the constituent gluon mass, \emph{}the infrared
regulator of the \HWPP\ parton shower.

Using this approach we can calculate the shower variables~$(\tilde{q}_h,z_h,\phi_z)$
for the hardest emission which allows us to generate the truncated and vetoed showers.

\section{Results\label{sec:Results}}
\begin{figure}
\begin{centering}\includegraphics[width=0.47\textwidth,angle=90]{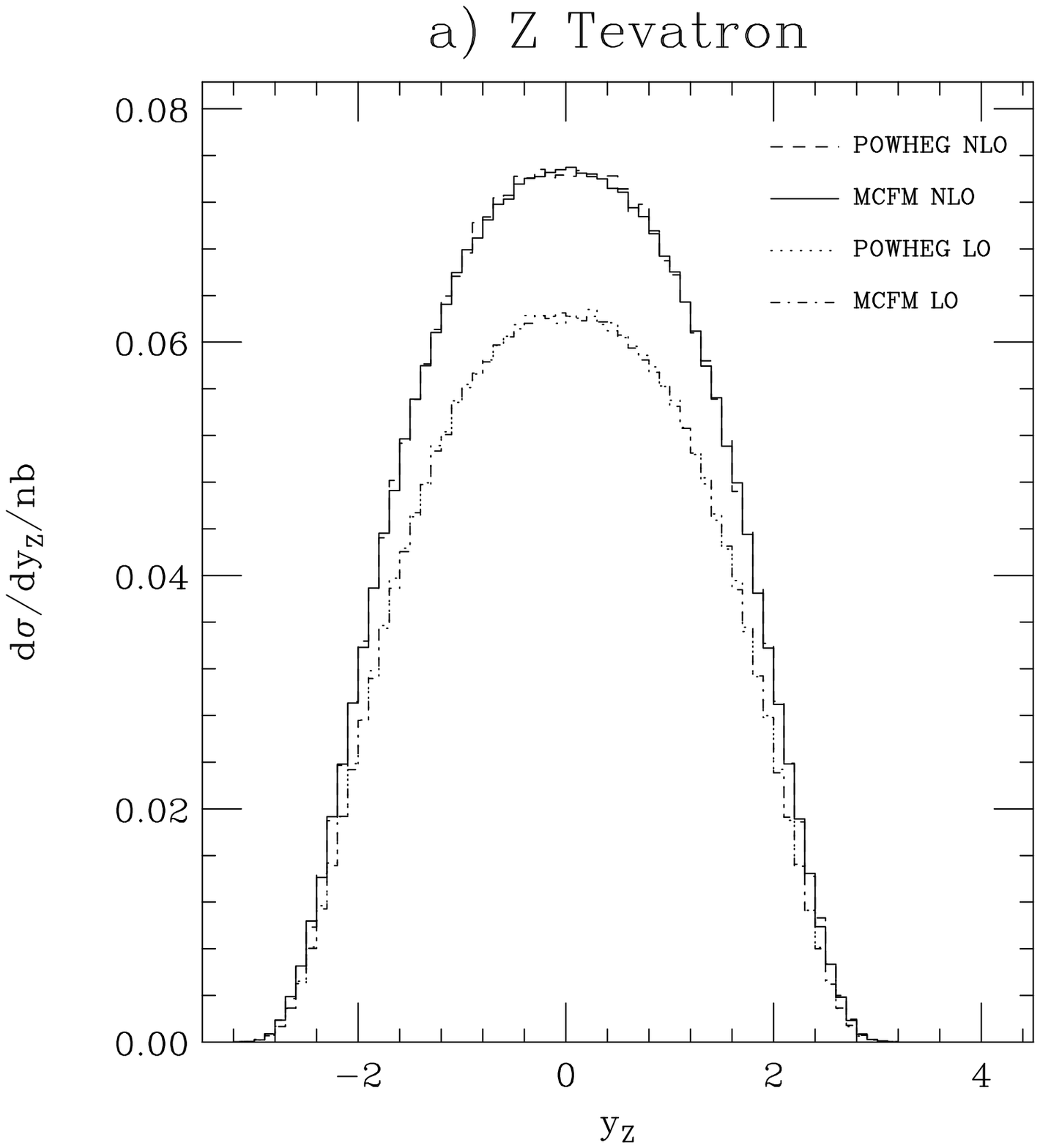}
\hfill{}\includegraphics[width=0.47\textwidth,angle=90]{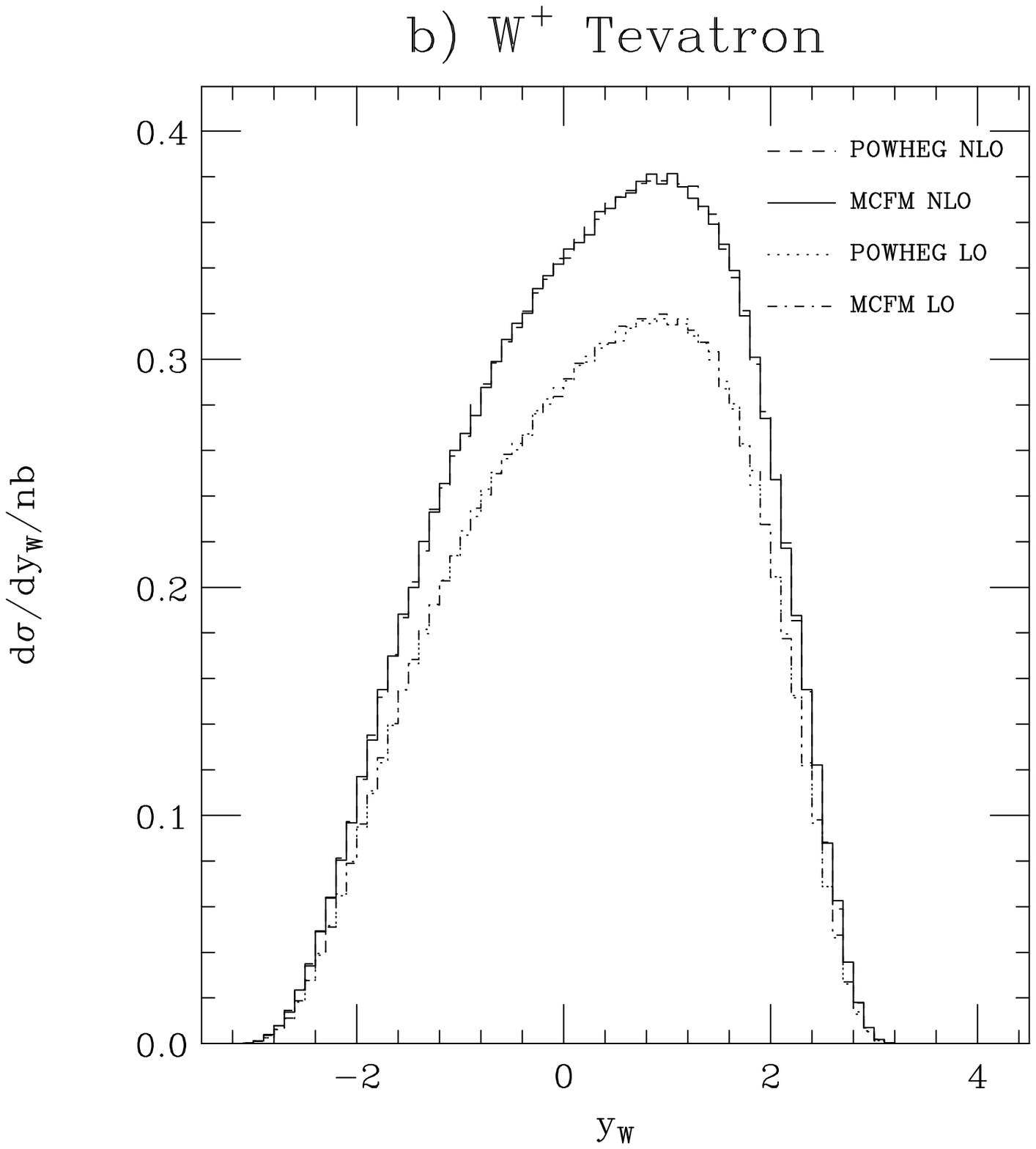}
\\
 \includegraphics[width=0.47\textwidth,angle=90]{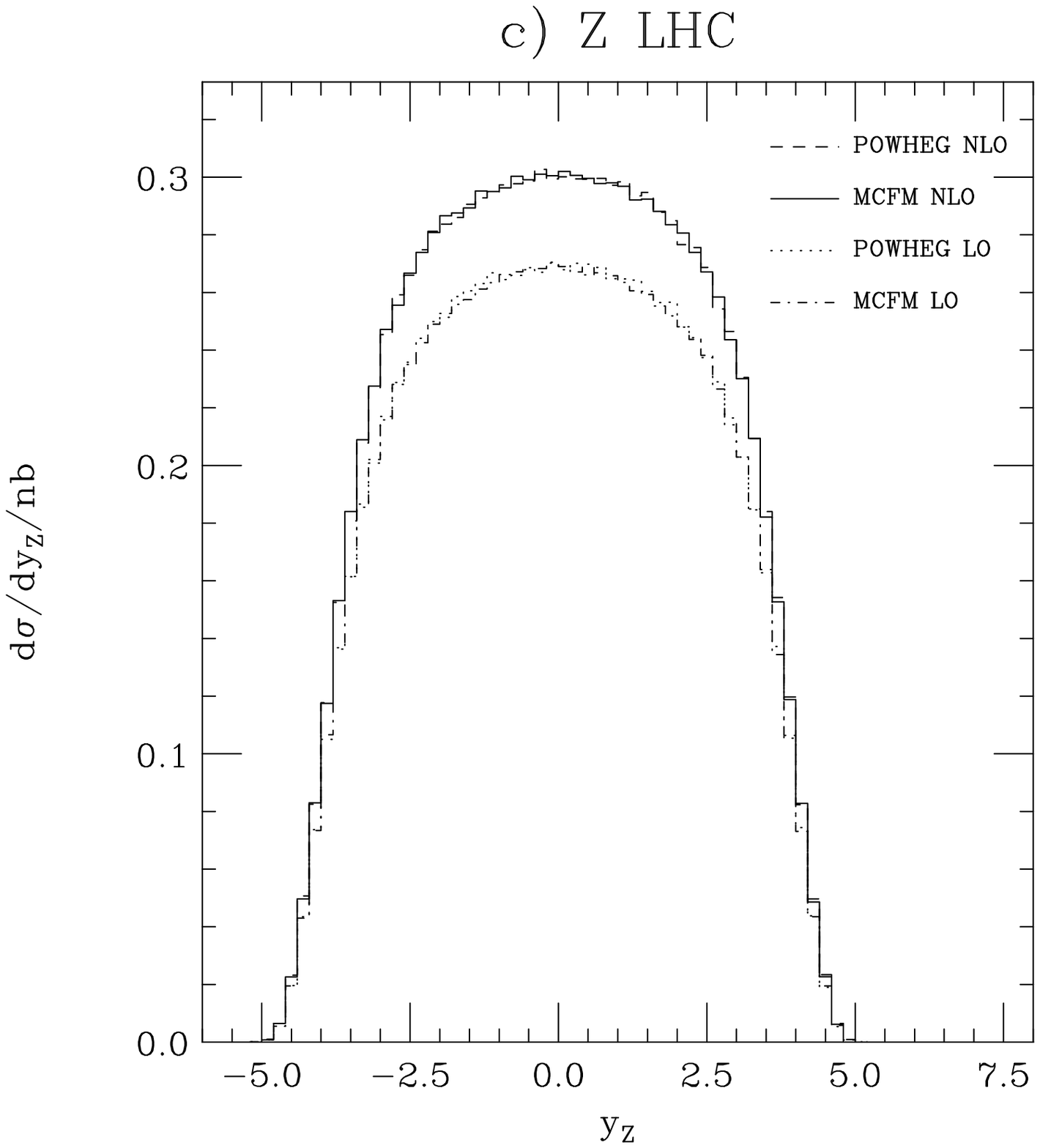} \hfill{}\includegraphics[width=0.47\textwidth,angle=90]{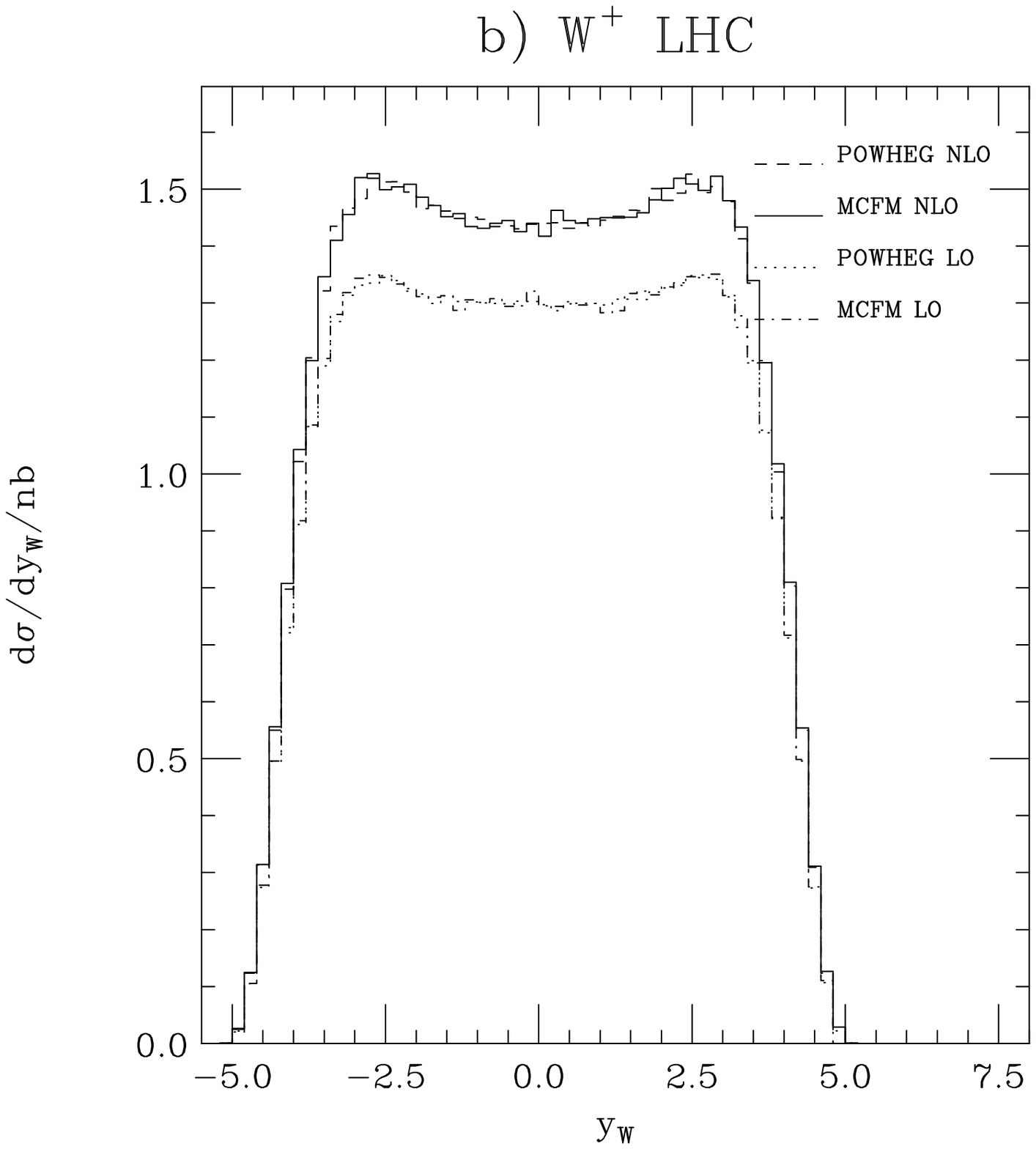}
\\
 \par\end{centering}

\caption{Comparisons of $\mathrm{d}\sigma/\mathrm{d}y$ for the \textsf{POWHEG}
implementation and \textsf{MCFM}~\cite{Campbell:2000bg} for $Z$ and $W^{+}$
production at the Tevatron ($\sqrt{s}=2\mathrm{TeV}$) and the LHC
($\sqrt{s}=14\mathrm{TeV}$).}
\label{MCFM} 
\end{figure}

\begin{figure}
\begin{center}
\includegraphics[angle=90,width=0.47\textwidth]{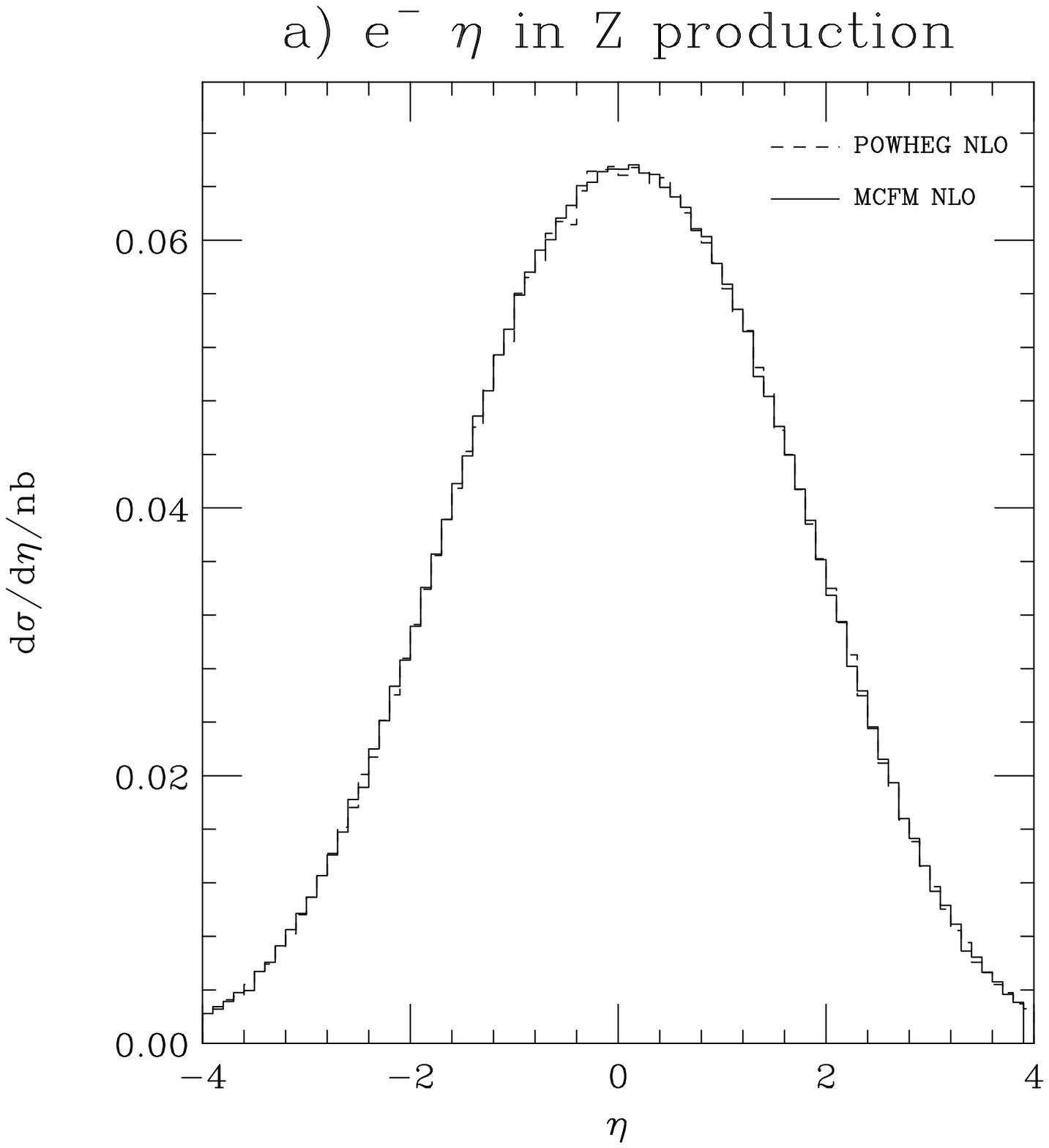}\hfill
\includegraphics[angle=90,width=0.47\textwidth]{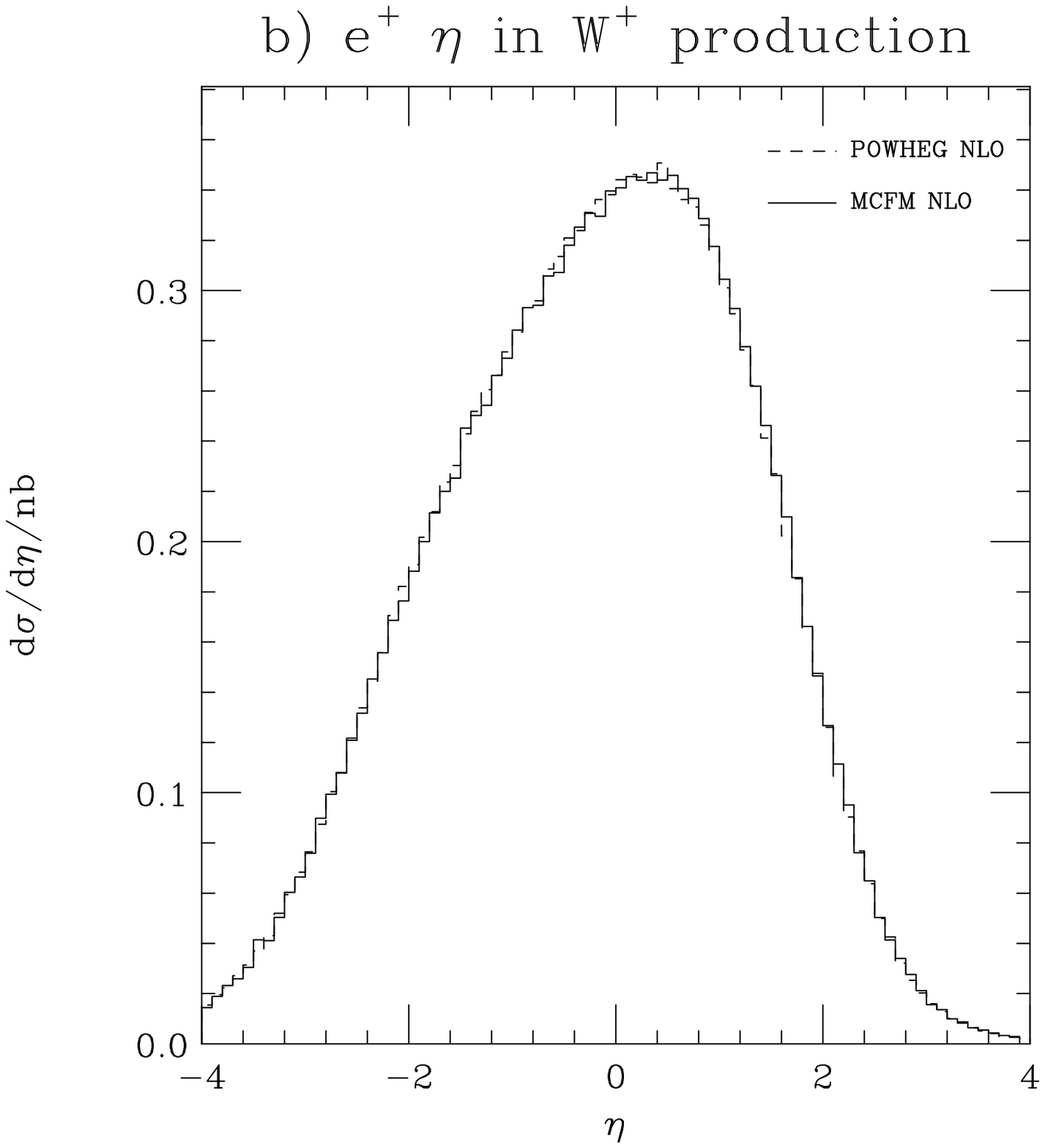}
\end{center}
\caption{The rapidity of a) the electron in $Z$ and b) the positron in $W^+$
     	production at the Tevatron including the leptonic decay of the gauge boson
     	for the \textsf{POWHEG} implementation and \textsf{MCFM}~\cite{Campbell:2000bg} at the
	Tevatron~($\sqrt{s}=2\mathrm{TeV}$).}
\label{fig:leptonrap}
\end{figure}

As a check of the calculation of the next-to-leading order differential cross section,
distributions of the vector boson rapidity produced by the
POWHEG implementation and the NLO program \textsf{MCFM}~\cite{Campbell:2000bg}
were compared. Fig.\,\ref{MCFM} shows distributions for $\gamma/Z$
and $W^{+}$ production at the Tevatron (proton-antiproton at $\sqrt{s}$\,=\,2\,TeV)
and the LHC (proton-proton at $\sqrt{s}$\,=\,14\,TeV). In all
cases the total cross sections from \textsf{MCFM} and our \textsf{POWHEG} implementation
agreed to within 0.5\,\%.
The distribution of the rapidity of the lepton produced in the $\gamma/Z$ and $W$ decay is shown in Fig.\,\ref{fig:leptonrap} and is also in good agreement.
 For both \HWPP\ and \textsf{MCFM} in this comparison,
the parton density functions used were the MRST2001 NLO \cite{Martin:2002dr}
set with the \textsf{LHAPDF} interface \cite{Whalley:2005nh}. 

\begin{figure}
\begin{centering}\includegraphics[width=0.61\textwidth,angle=90]{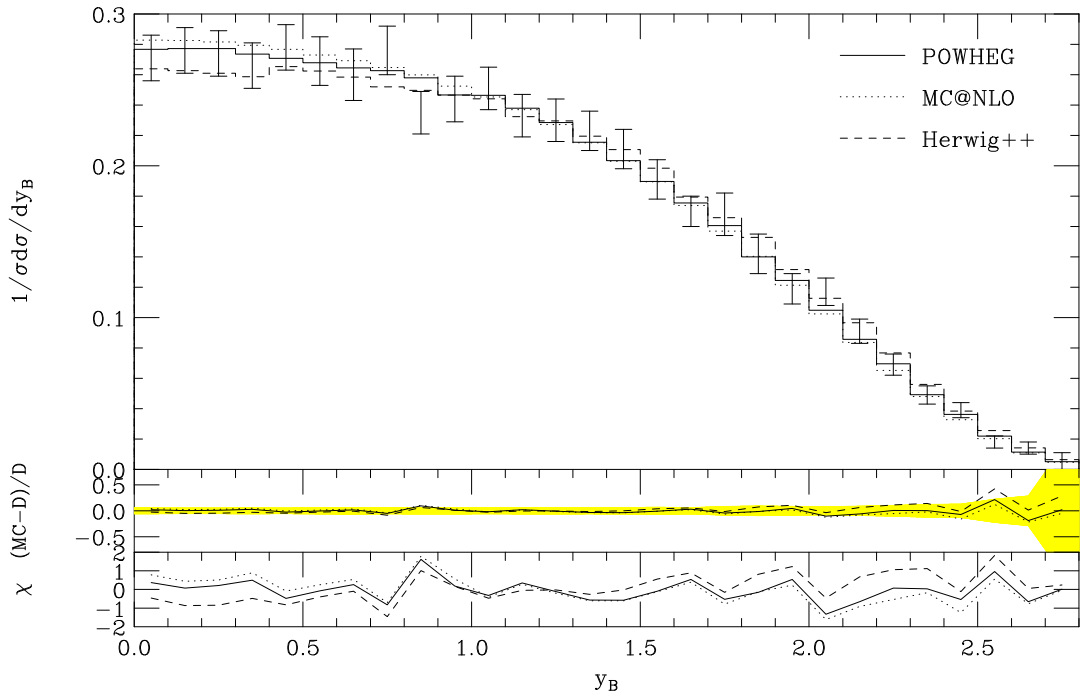} \par\end{centering}
\caption{Rapidity distribution for $Z$ production compared to D0 Run II Tevatron
data~\cite{Abazov:2007jy}. The solid line shows the prediction of
our \textsf{POWHEG} implementation, the dotted line is the prediction
of \textsf{MC@NLO} and the dashed line is the default \HWPP\ result. }

\label{Zrap} 
\end{figure}

\begin{figure}
\begin{centering}\includegraphics[width=0.65\textwidth,angle=90]{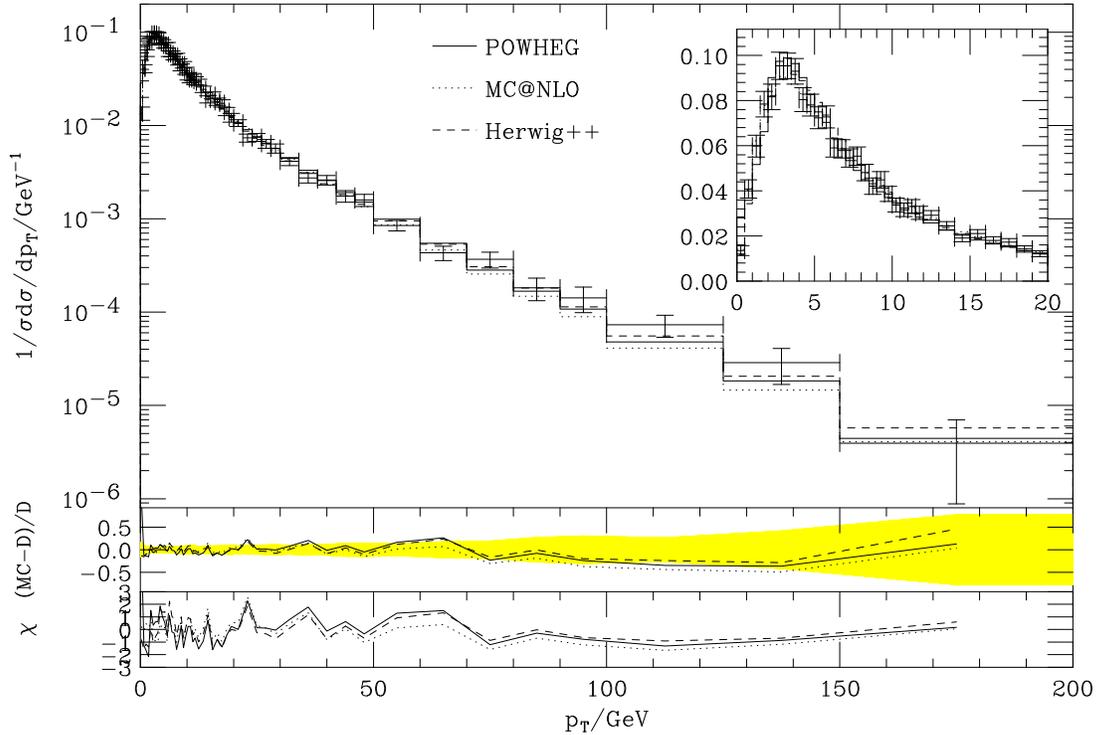} \par\end{centering}
\caption{Transverse momentum distribution for $Z$ production compared to
CDF Run I Tevatron data \cite{Affolder:1999jh}. The solid line shows
the prediction of our \textsf{POWHEG} implementation, the dotted line
is the prediction of \textsf{MC@NLO} and the dashed line is the default
\HWPP\ result. The inset shows an expanded view of the low $p_{T}$
region.}

\label{Zpt1} 
\end{figure}

\begin{figure}
\begin{centering}\includegraphics[width=0.67\textwidth,angle=90]{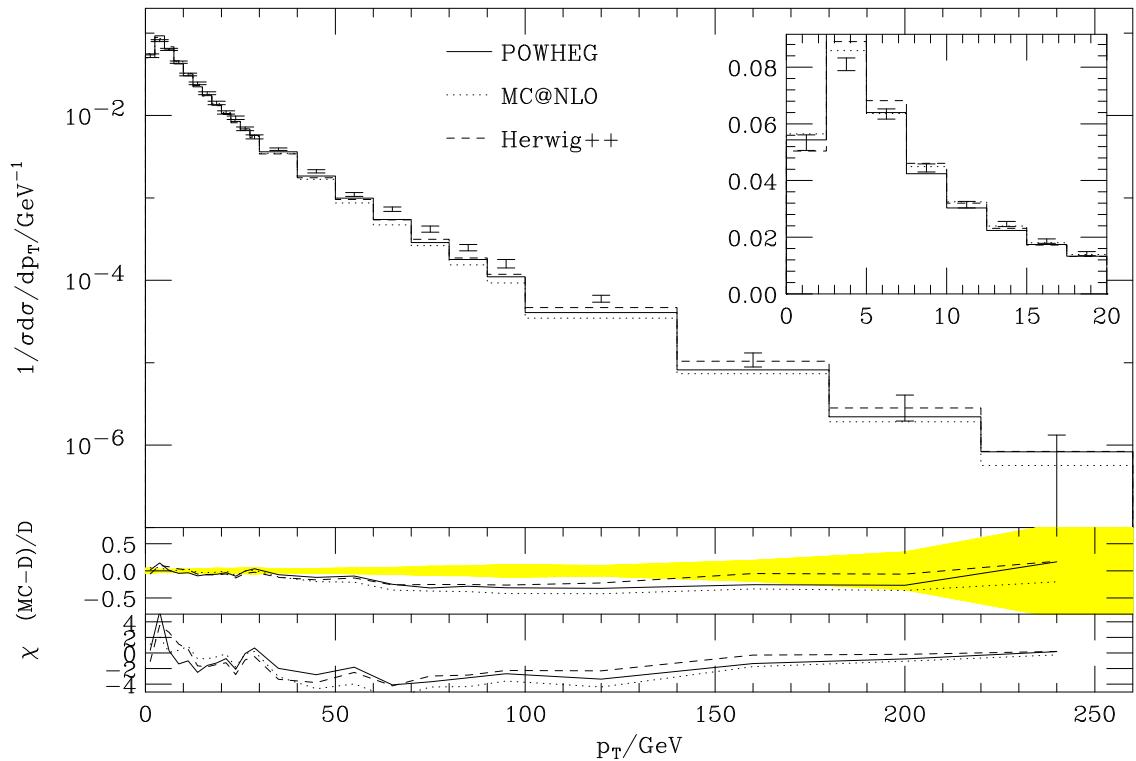} \par\end{centering}
\caption{Transverse momentum distribution for $Z$ production compared to
D0 Run II Tevatron data \cite{Abazov:2007nt}. The solid line shows
the prediction of our \textsf{POWHEG} implementation, the dotted line
is the prediction of \textsf{MC@NLO} and the dashed line is the default
\HWPP\ result. The inset shows an expanded view of the low $p_{T}$
region.}

\label{Zpt2} 
\end{figure}

\begin{figure}
\begin{centering}\includegraphics[width=0.67\textwidth,angle=90]{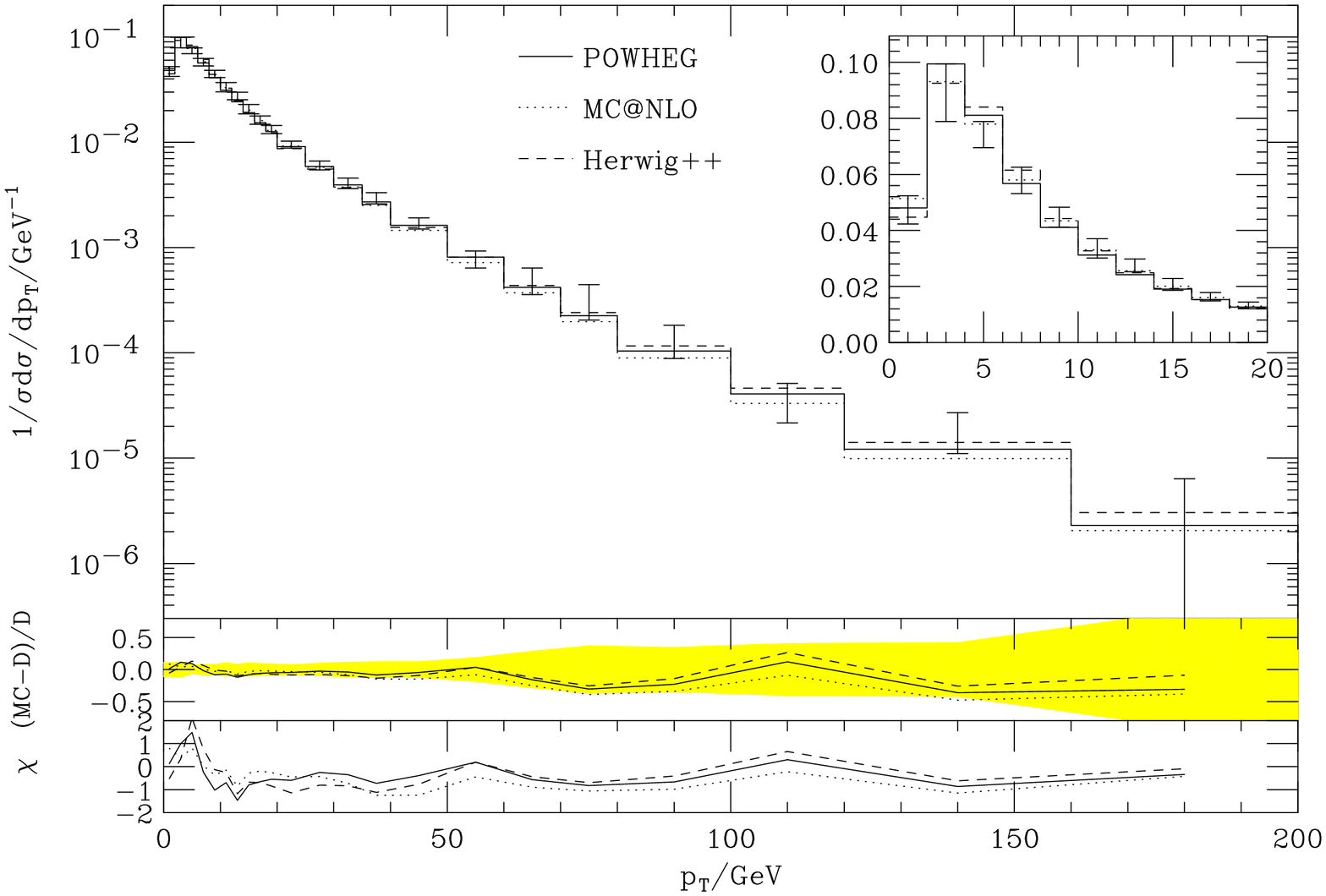} \par\end{centering}

\caption{Transverse Momentum distribution for $W$ production compared to
D0 Run I data \cite{Abbott:1998jy}. The solid line shows the prediction
of our \textsf{POWHEG} implementation, the dotted line is the prediction
of \textsf{MC@NLO} and the dashed line is the default \HWPP\ result.
The inset shows an expanded view of the low $p_{T}$ region.}

\label{Wpt} 
\end{figure}

\begin{figure}
\begin{center}
\includegraphics[angle=90,width=0.49\textwidth]{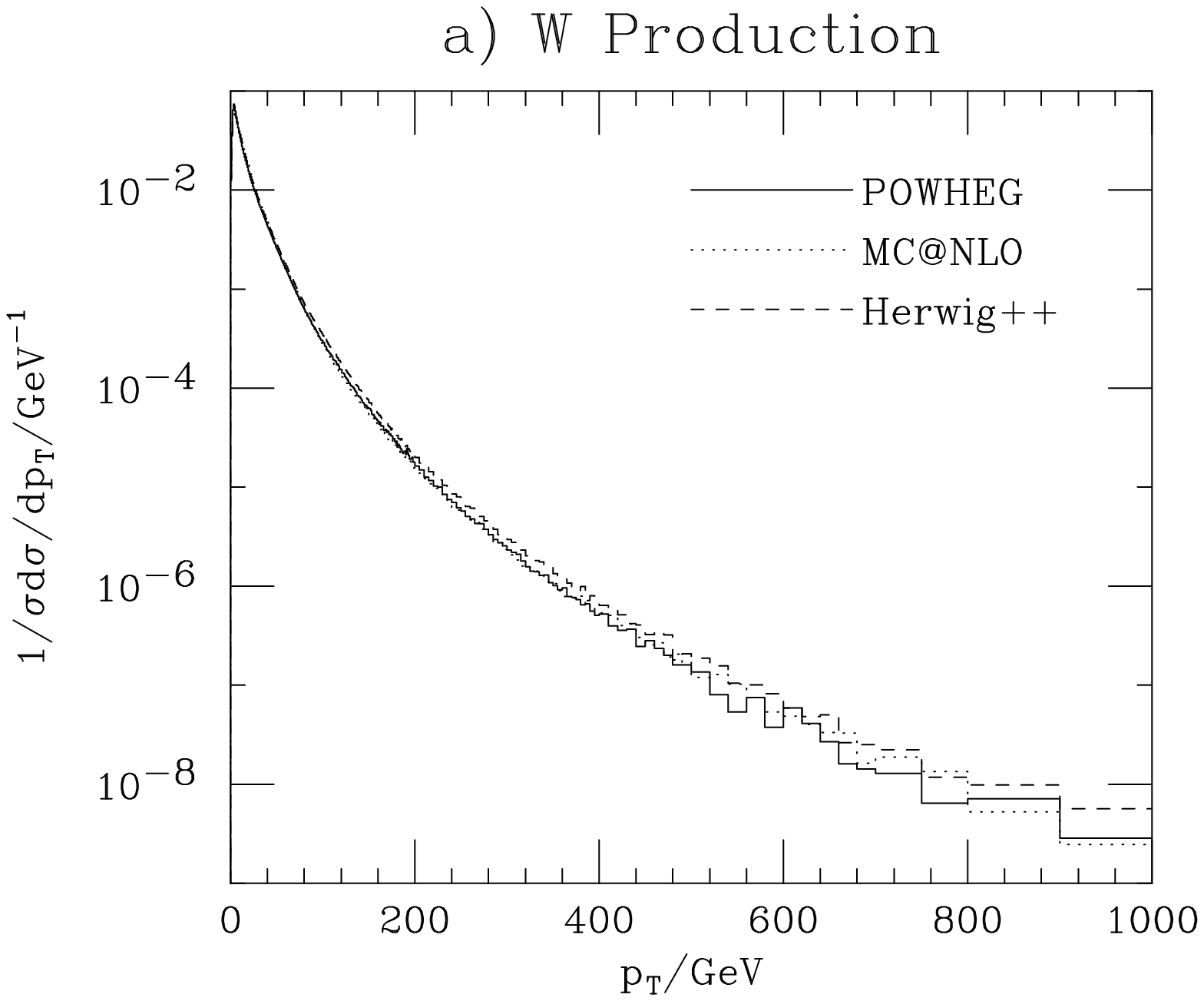}\hfill
\includegraphics[angle=90,width=0.49\textwidth]{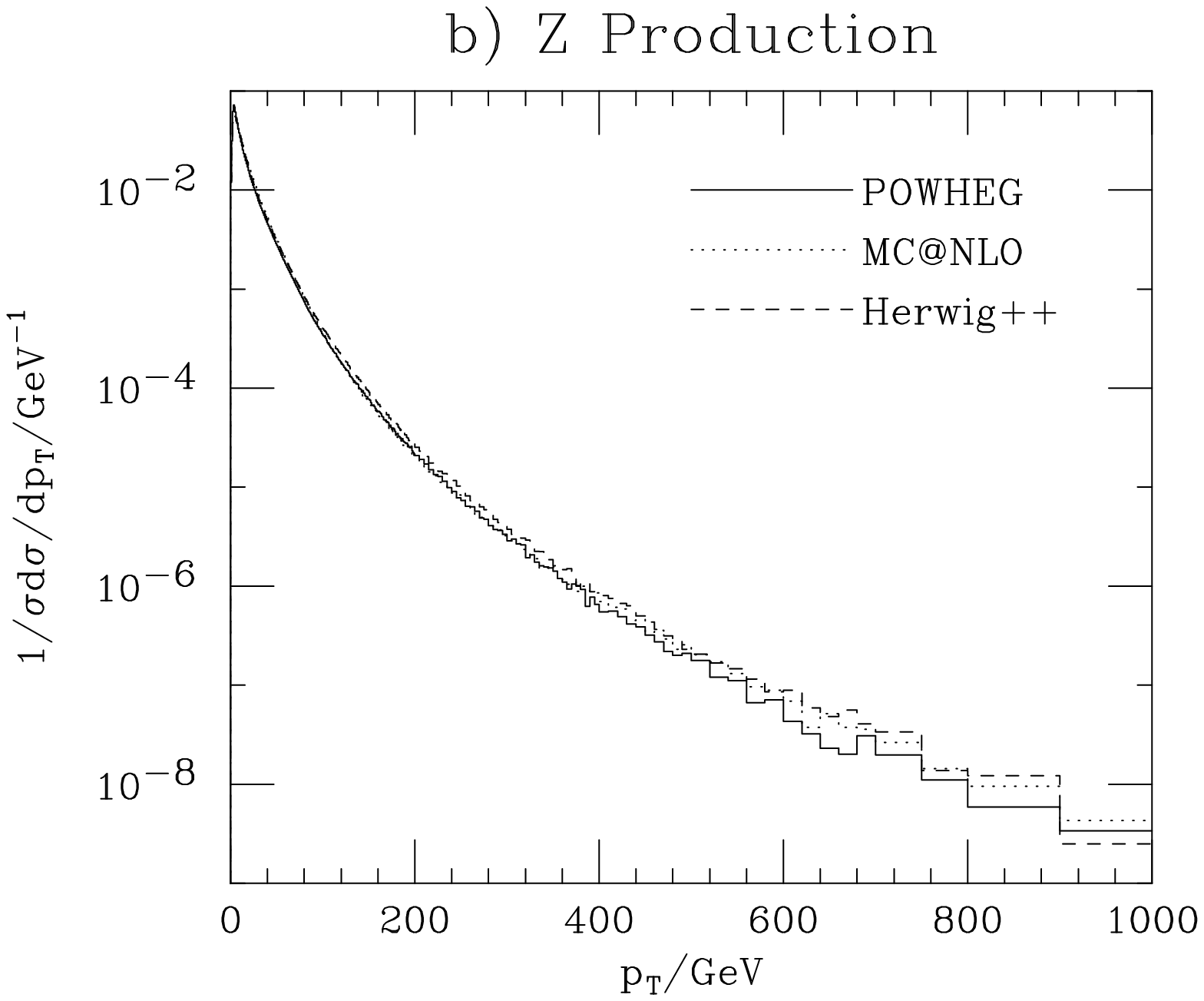}
\end{center}
\vspace{-4mm}
\caption{The $p_T$ distributions of a) $W$ and b) $Z$ bosons at the LHC. 
	The solid line shows the prediction
	of our \textsf{POWHEG} implementation, the dotted line is the prediction
	of \textsf{MC@NLO} and the dashed line is the default \HWPP\ result.}
\label{fig:LHC}
\end{figure}

In Figs.\,\ref{Zrap}-\ref{Wpt}, distributions from the Drell-Yan
POWHEG implementations for the rapidity and transverse momentum of
the vector bosons are compared to Tevatron data. The middle and bottom 
panels in each of these plots shows the $(\mathrm{Theory}-\mathrm{Data})/\mathrm{Data}$
and $\chi$
values for each bin. In Fig. \ref{Zrap} the rapidity distribution
of $\gamma/Z$ bosons of mass 71-111\,GeV is compared to D0 Run II
data \cite{Abazov:2007jy}. Fig. \ref{Zpt1} shows the transverse
momentum distribution of $\gamma/Z$ bosons of mass 66-116\,GeV compared
to CDF Run I data \cite{Affolder:1999jh}. Fig. \ref{Zpt2} shows
the transverse momentum distribution of $\gamma/Z$ bosons of mass
40-200\,GeV compared to D0 Run II data \cite{Abazov:2007nt}. Fig.
\ref{Wpt} shows the transverse momentum distribution of $W$
bosons compared to Run I D0 data \cite{Abbott:1998jy}. In addition
to the results from our implementation of the \textsf{POWHEG} method
the results from \HWPP\ including a matrix element correction
and \textsf{MC@NLO}~\cite{Frixione:2002ik,Frixione:2003ei,Frixione:2005vw,Frixione:2006gn,Frixione:2007zp,Frixione:2008yi}
are shown. The predicted $W$ and $Z$ $p_T$ distributions at the LHC
are shown in Fig.\,\ref{fig:LHC}.

The \HWPP\ results were generated using an intrinsic $p_{T}$ of
2.2\,GeV which was obtained by fitting to the Run I $W$ and $Z$
$p_{T}$ distributions~\cite{Bahr:2008pv}. The \textsf{POWHEG} results
used the same intrinsic $p_{T}$ as for \HWPP\ and a minimum $p_{T}$
of 2\,GeV for the hardest emission. The \textsf{MC@NLO} and \textsf{HERWIG}
results were generated using an intrinsic $p_T$ of 1.6\,GeV from
a fit to D0 data~\cite{Nurse:2005vh}.

The leading-order parton distribution
functions of \cite{Martin:2002dr} were used for the \HWPP\ result
and the central value of the NLO parton distributions of \cite{Martin:2002aw}
for the \textsf{POWHEG} and \textsf{MC@NLO} results.

All the approaches give good agreement for the rapidity of the $Z$
boson however they differ in the description of the $p_{T}$ spectrum
of the gauge boson. The chi squared per degree of freedom for the
various $p_{T}$ spectra and approaches are given in Table~\ref{tab:chisq}.
All the approaches are in good agreement with the Run I data from
CDF and D0 for the $p_{T}$ of the $Z$ and $W$. However, with the
exception of the results of the \textsf{FORTRAN} \textsf{HERWIG} program
including a matrix element correction, which gave the worst agreement
with the Run I $Z$ data, all the results are below the new D0 $Z$
$p_{T}$ data at high transverse momentum.

There is a common trend, at both the Tevatron and LHC energies,
that the matrix element correction gives the
largest result at large $p_{T}$, followed by the \textsf{POWHEG}
approach with \textsf{MC@NLO} giving the lowest value. This is due
to the treatment of the hardest emission in the different approaches.
In the \textsf{MC@NLO} method the result at large $p_{T}$ is the leading-order
matrix element for the production of a vector boson and a hard
QCD jet. However in this region, as we are normalising to the total
cross section, the matrix element correction result is essentially
the matrix element for vector boson plus jet production multiplied
by the K-factor%
\footnote{The K-factor here is the ratio of the NLO cross section for inclusive
vector boson production divided by the leading-order cross section.%
} giving a larger result. In the large $p_{T}$ region the \textsf{POWHEG}
result, because the real-emission matrix element is exponentiated,
is the real-emission matrix element multiplied by the $\bar{B}$ function,
which results in a K-factor-like correction, and the Sudakov form
factor which causes the result to be slightly smaller than the default~\HWPP\
 result. The \textsf{POWHEG} result has the significant advantage
that rather than using a global rescaling of the cross section to
get the NLO normalization, which can lead to a poor description of
observables, such as the boson rapidity, which are non-zero at leading
order the NLO correction is calculated for each momentum configuration.

In an ideal world we would like to use the NLO result for vector boson
production in association with a hard jet to describe the high $p_{T}$
region, however incorporating this result into a Monte Carlo simulation
is not currently feasible, and therefore the \textsf{POWHEG} or matrix
element correction methods which basically assume the correction for
inclusive production is the same as for vector boson production in
association with a jet at least have the advantage of including a
correction which improves agreement with data.

\begin{table}
\begin{centering}\begin{tabular}{|c|c|c|c|c|c|c|}
\hline 
Approach &
\multicolumn{6}{|c|}{Data Set}\tabularnewline
\hline 
&
\multicolumn{2}{|c|}{D0 W $p_{T}$}&
\multicolumn{2}{|c|}{CDF Z $p_{T}$}&
\multicolumn{2}{|c|}{D0 Z $p_{T}$}\tabularnewline
\hline 
&
All&
$p_{T}>30$\,GeV &
All&
$p_{T}>30$\,GeV &
All &
$p_{T}>30$\,GeV \tabularnewline
\hline 
\textsf{MC@NLO} &
0.51 &
0.82 &
0.70 &
0.96 &
7.2 &
13.9\tabularnewline
\HWPP\  &
0.67 &
0.42 &
0.89 &
0.61 &
5.1 &
7.0\tabularnewline
\textsf{POWHEG} &
0.54 &
0.33 &
1.99 &
1.00 &
5.3 &
6.9\tabularnewline
\textsf{HERWIG} &
0.69 &
1.08 &
2.45 &
4.47 &
2.0 &
1.9\tabularnewline
\hline
\end{tabular}\par\end{centering}
\caption{Chi squared per degree of freedom for \textsf{MC@NLO}, \HWPP, our
implementation of the \textsf{POWHEG} method in \HWPP\ and \textsf{FORTRAN}
\textsf{HERWIG} compared to Tevatron vector boson $p_{T}$ data. The chi-squared values are calculated for the shapes of the distributions, \emph{i.e.} normalizing them to unity. In
order to compare the high $p_{T}$ region and minimise the effect
of tuning the intrinsic transverse momentum the chi squared per degree
of freedom is given for both the full $p_{T}$ region and only for
the data points with $p_{T}>30$\,GeV.}

\label{tab:chisq} 
\end{table}

\begin{figure}
\begin{center}
\includegraphics[width=0.67\textwidth,angle=90]{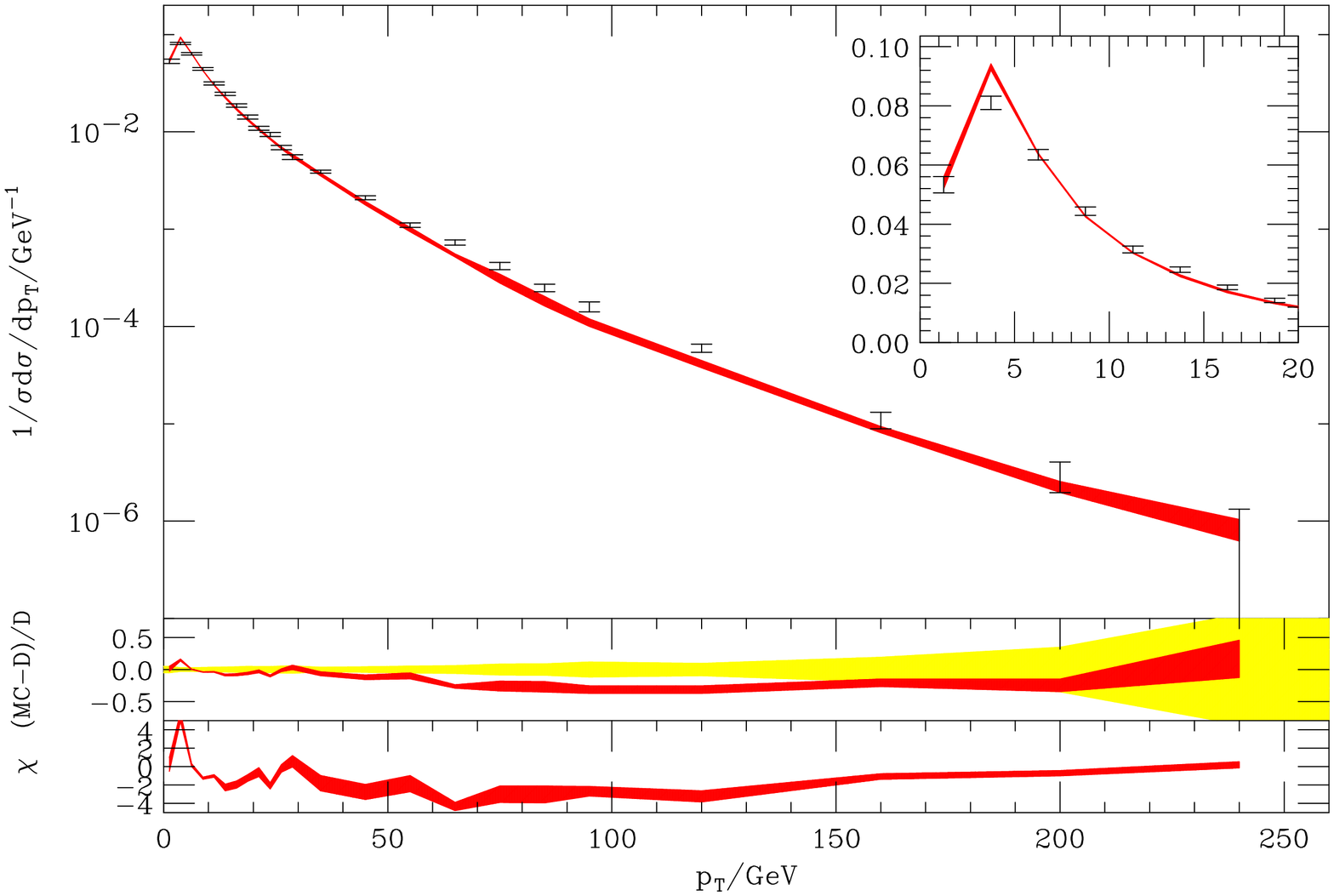}
\end{center}
\vspace{-3mm}
\caption{Transverse Momentum distribution for $Z$ production compared to
D0 Run II data \cite{Abazov:2007nt}.  The band shows the effect of varying the scale
used for the parton distributions and $\alpha_S$
between $0.5\hat{s}$ and $2\hat{s}$ for the $\overline{B}$ term and
between $0.5(M_B^2+p_T^2)$ and $2(M_B^2+p_T^2)$ for the hardest emission.}
\label{fig:scale}
\end{figure}

In general all the results lie below the D0 Run II $Z$ $p_{T}$ data
between 50~and 100\,GeV which results in the relatively poor chi
squared, however in general the \textsf{POWHEG} approach gives comparable
results to the other state-of-the-art techniques. The effect of varying the scale
used for the parton distributions and $\alpha_S$
between $0.5\hat{s}$ and $2\hat{s}$ for the $\overline{B}$ term and
between $0.5(M_B^2+p_T^2)$ and $2(M_B^2+p_T^2)$ for the hardest emission
is shown in Fig.\,\ref{fig:scale}. While this variation moves the POWHEG result close to the data it still is below the experimental result in the intermediate $p_T$ region.

\begin{figure}
\begin{center}
\includegraphics[width=0.49\textwidth,angle=90]{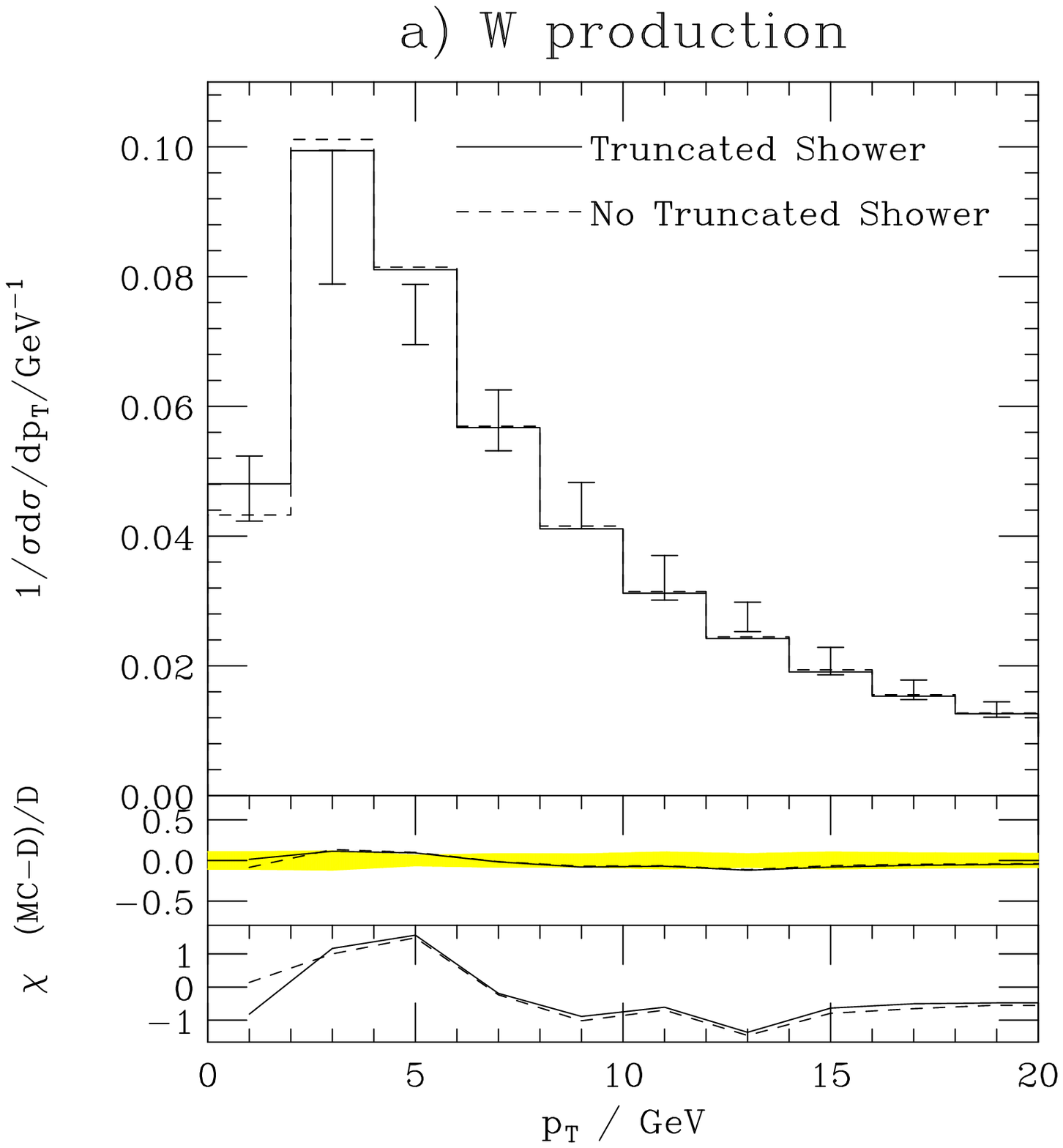}\hfill
\includegraphics[width=0.49\textwidth,angle=90]{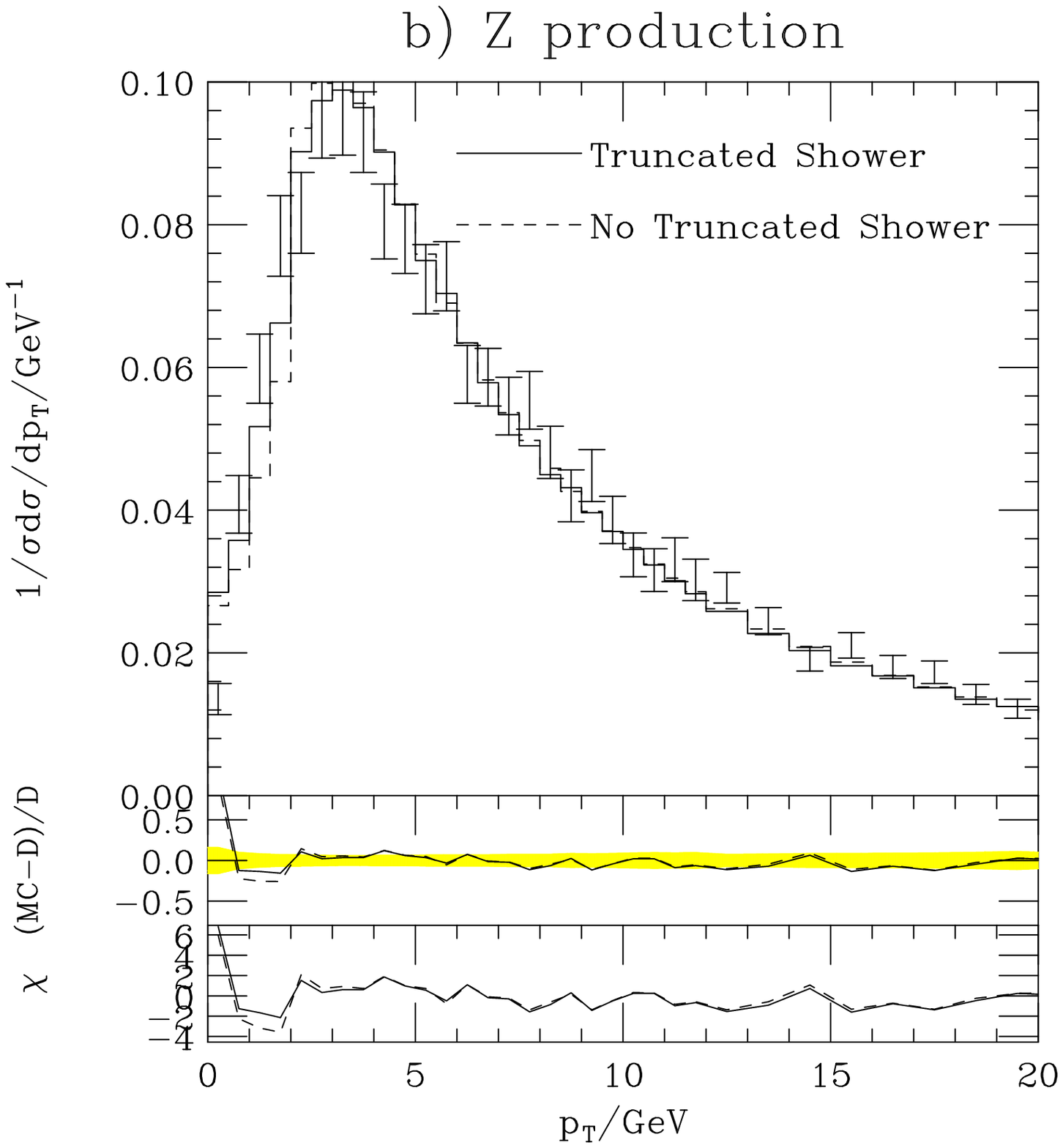}
\end{center}
\vspace{-3mm}
\caption{Transverse Momentum distribution for a) $W$ production compared to
D0 Run I data \cite{Abbott:1998jy} and b) $Z$ production compared to
CDF Run I Tevatron data \cite{Affolder:1999jh}.
The solid line includes the truncated shower whereas the dashed line does not.}
\label{fig:truncated}
\end{figure}

  The effect of the truncated shower is illustrated in Fig.\,\ref{fig:truncated}
which shows the small $p_T$ region of the transverse momentum distribution for
$W$ and $Z$ production compared to D0 and CDF data. In this region where the highest
$p_T$ emission is at a small scale and there is often a large region for the evolution
of the truncated shower it has the largest effect. However the effect is
relatively small at least for the transverse momentum distribution, equivalent
to a small change in the intrinsic transverse momentum. We would expect to see a larger
effect in the distributions of jets in the event, in particular the second hardest jet, 
which we will study in the future.

\section{Conclusion\label{sec:Conclusion}}

The \textsf{POWHEG} NLO matching prescription has been implemented
in the \HWPP\ Monte Carlo event generator for Drell-Yan vector boson
production. A full treatment of the truncated shower, which is required
to produce wide angle, soft radiation in angular-ordered parton showers,
is included for the first time.

The implementation gives a good description of data, on a similar
level to the matrix element correction methods and better than \textsf{MC@NLO}. It
will be available in a forthcoming release of \HWPP.

The technique we have used to implement the \textsf{POWHEG} approach,
by interpreting the hard emission in terms of the variables used to
generate the parton shower is very powerful, and has many other applications
in matching approaches, for example for the 
CKKW approach~\cite{Catani:2001cc,Krauss:2002up},
which we will explore in the future.

\acknowledgments

We are grateful to the other members of the \HWPP\ collaboration 
and Paolo Nason for many useful discussions. We are particularly 
grateful to Mike Seymour for his careful reading of this manuscript.
This work was supported by the Science
and Technology Facilities Council, formerly the Particle Physics and
Astronomy Research Council, the European Union Marie Curie Research
Training Network MCnet under contract MRTN-CT-2006-035606. K. Hamilton
acknowledges support from the Belgian Interuniversity Attraction Pole,
PAI, P6/11.
\appendix

\section*{Note added in Proof}

  While we were in the final stages of completing this paper another paper on
  the same topic was submitted to the arXiv~\cite{Nason:2008new}.

\section{Plus Distributions\label{sec:Plus-distributions}}

In order to implement the collinear $\left(\mathcal{C}_{ab}\right)$
terms in the real-emission contributions to $\overline{B}\left(\Phi_{B}\right)$
the following relations are required
\begin{equation}
\begin{array}{rcl}
\int_{\bar{x}\left(v\right)}^{1}\mathrm{d}x\textrm{ }\frac{f\left(x\right)}{\left(1-x\right)_{+}} & = & \int_{0}^{1}\mathrm{d}\tilde{x}\textrm{ }\left(1-\bar{x}\left(v\right)\right)\left[\frac{f\left(x\left(\tilde{x},v\right)\right)-f\left(x\left(1,v\right)\right)}{1-x\left(\tilde{x},v\right)}+\frac{f\left(x\left(1,v\right)\right)}{1-\bar{x}\left(v\right)}\ln\left(1-\bar{x}\left(v\right)\right)\right]\end{array}\label{eq:plus_fns_1}\end{equation}
and\begin{equation}
\begin{array}{ll}
 & \int_{\bar{x}\left(v\right)}^{1}\mathrm{d}x\textrm{ }f\left(x\right)\left(\frac{\ln\left(1-x\right)}{1-x}\right)_{+}\\
= & \int_{0}^{1}\mathrm{d}\tilde{x}\textrm{ }\left(1-\bar{x}\left(v\right)\right)\left[\left(f\left(x\left(\tilde{x},v\right)\right)-f\left(x\left(1,v\right)\right)\right)\left(\frac{\ln\left(1-x\left(\tilde{x},v\right)\right)}{1-x\left(\tilde{x},v\right)}\right)+\frac{f\left(x\left(1,v\right)\right)}{2\left(1-\bar{x}\left(v\right)\right)}\ln^{2}\left(1-\bar{x}\left(v\right)\right)\right]\end{array}\label{eq:plus_fns_2}\end{equation}
with $\tilde{x}$ defined in Eq.\,\ref{eq:impl_1} and $v\in\left[0,1\right]$.
For the hard $\left(\mathcal{H}_{ab}\right)$ contribution to the
real radiation components in $\overline{B}\left(\Phi_{B}\right)$
\begin{equation}
\begin{array}{ll}
 & \int_{0}^{1}\mathrm{d}v\int_{\bar{x}\left(v\right)}^{1}\mathrm{d}x\mbox{ }f\left(x,v\right)\mbox{ }\frac{1}{\left(1-x\right)_{+}}\left(\frac{1}{\left(1-v\right)_{+}}+\frac{1}{v_{+}}\right)\\
= & \int_{0}^{1}\mathrm{d}v\int_{0}^{1}\mathrm{d}\tilde{x}\mbox{ }\frac{1}{1-\tilde{x}}\left(\frac{f\left(x\left(\tilde{x},v\right),v\right)-f\left(1,v\right)-f\left(x\left(\tilde{x},1\right),1\right)+f\left(1,1\right)}{1-v}+\frac{f\left(x\left(\tilde{x},v\right),v\right)-f\left(1,v\right)-f\left(x\left(\tilde{x},0\right),0\right)+f\left(1,0\right)}{v}\right)\\
+ & \int_{0}^{1}\mathrm{d}v\int_{0}^{1}\mathrm{d}\tilde{x}\mbox{ }\left(\frac{f\left(1,v\right)\ln\left(1-\bar{x}\left(v\right)\right)-f\left(1,1\right)\ln\left(1-\bar{x}\left(1\right)\right)}{1-v}+\frac{f\left(1,v\right)\ln\left(1-\bar{x}\left(v\right)\right)-f\left(1,0\right)\ln\left(1-\bar{x}\left(0\right)\right)}{v}\right),\end{array}\label{eq:plus_fns_3}\end{equation}
where in the last line of Eq.\,\ref{eq:plus_fns_3}
we have introduced the identity as $\int_{0}^{1}\mathrm{d}\tilde{x}$.
Similar relations are derived, in different variables, in Ref.\,\cite{Frixione:2007vw}.

\providecommand{\href}[2]{#2}\begingroup\raggedright\endgroup
\end{document}